\documentclass[reprint,twocolumn,amsmath,amssymb,superscriptaddress,nofootinbib]{revtex4-2}

\usepackage[english]{babel}

\usepackage[svgnames]{xcolor}
\usepackage{graphicx, graphpap}
\usepackage[caption=false,subrefformat=parens,labelformat=parens]{subfig}
\usepackage{amsmath}
\usepackage{physics} 
\usepackage{comment} 
\usepackage{multirow}
\usepackage[T1]{fontenc}
\usepackage{bbold} 
\usepackage{mathtools}
\usepackage{footmisc}
\usepackage[normalem]{ulem}
\usepackage{xurl}
\usepackage[colorlinks=true, allcolors=blue]{hyperref}

\allowdisplaybreaks

\begin{document}
\title{Temporal trade-offs in high-dimensional entanglement: a comprehensive noise model for optimal time-bin QKD protocols}

\author{Alexandra E. Bergmayr-Mann}
\email{alexandra.bergmayr@gmx.at}
\affiliation{TU Wien, Atominstitut \&  Vienna Center for Quantum Science and Technology, Stadionallee 2, 1020 Vienna, Austria}
\author{Gláucia Murta}
\email{glaucia.murta@tuwien.ac.at}
\affiliation{TU Wien, Atominstitut \&  Vienna Center for Quantum Science and Technology, Stadionallee 2, 1020 Vienna, Austria}
\author{Marcus Huber}
\email{marcus.huber@tuwien.ac.at}
\affiliation{TU Wien, Atominstitut \&  Vienna Center for Quantum Science and Technology, Stadionallee 2, 1020 Vienna, Austria}
 \affiliation{Institute for Quantum Optics and Quantum Information (IQOQI),
Austrian Academy of Sciences, Boltzmanngasse 3, 1090 Vienna, Austria}
\date{\today}

\begin{abstract}
High-dimensional time-bin entanglement is known for having high potential for quantum key distribution (QKD) applications, being easily implementable, robust and may offer better keyrates than simple qubit protocols. The temporal encoding, combined with limited clock resolution, however, implies a trade-off: \textit{Is it better to send more low-dimensionally encoded photons or few high-dimensional ones?} Answering this question presents a hard challenge as it depends on the entire context of the protocol, including the production rates, losses, dark counts, timing jitters and many more. We answer this question for single Franson based interferometers and asymptotic key rate and shared entanglement as main figures of merit. We present a flexible noise model incorporating all relevant parameters and find that there are indeed regions where high-dimensional encoding still outperforms any qubit based version when the pair arrival rate is limited, for instance through loss or limited pump power.
\end{abstract}

\maketitle

\section{Introduction}
\subsection{Background}
Quantum Key Distribution (QKD) enables two remote parties to establish secure keys, resilient even against eavesdroppers with unlimited computational power. The resource underlying this quantum physical application is quantum entanglement \cite{Curty_2004}, one of the defining features of quantum mechanics, which can be quantified by metrics such as the Entanglement of Formation (EoF) and the fidelity to a maximally entangled state.

A promising approach to QKD implementations is the realization via satellite-based QKD, overcoming the exponential loss experienced in optical fibers by utilizing free-space links for which loss scales only quadratically with the distance. 
However, satellite-earth links are naturally vulnerable to outside noise such as environmental photons, causing severe limitations on the practicality of these systems. High-Dimensional (HD) entanglement has been shown to enhance background noise-resistance, with HD entanglement in the time-domain being one of the most promising realizations for free-space applications. The feasibility and advantages of HD entanglement in entanglement distribution tasks have already been demonstrated \cite{Cozzolino_2019, Ecker_2019, Yu_2025, Bulla_2023_13, Doda_2021, Bulla_2023_107}.
However, in practical experiments, the actual advantage of high-dimensional states over qubits depends on the specific physical parameters of the implementation. Consequently, it is unclear whether a two-dimensional implementation might have outperformed a high-dimensional one under a different choice of tunable parameters. Given that experimental prototyping is highly resource-intensive and expensive, an exhaustive empirical search for optimal configurations is infeasible. Therefore, a representative theoretical model is needed to resolve that uncertainty and to additionally enable the determination of the best dimension and experimental parameter constellation for implementations a priori. 

The parameters are of course dependent on the actual implementation of the high-dimensional Hilbert space, where common approaches include spatial modes \cite{Scarfe_2026}, orbital angular momentum \cite{Zhang_2026, Kysela_2020}, frequency modes \cite{Tagliavacche_2025, Kashi_2025, Chang_2026}, temporal modes \cite{Brecht_2015} and discrete time-bins \cite{Euler_2025, Yu_2025, White_2025, Vagniluca_2020}. As the latter are suitable for both free-space and fiber transmission, and most commonly employed, we focused the noise model on this degree, which also exhibits the most intricate trade-off between dimensionality and pair-production rate.

In this work, we introduce a refined model which includes all typical sources of implementation noise, as well as detector jitter, demonstrating that there are indeed realistic scenarios in which HD entanglement prevails, as evidenced by superior key rates and metrics of entanglement.

Furthermore, we provide a rigorous formal justification for the density matrix reconstruction - which serves as the basis for calculating the secret key rate - as well as for the specific lower-bounding techniques applied for the EoF and the fidelity \cite{Martin_2017, Terhal_2000}. This establishes the mathematical soundness of our techniques and the reliability of the resulting bounds.

\subsection{Outline of the modeling technique}
The aim of our noise- and jitter- model is to reliably bound the elements of the density matrix representing the time-entangled state responsible for the observed coincidence clicks, which in turn can be used for lower-bounding the key rate.

To this end, after the description of the experimental setup and procedure in Section \ref{sec:setup} and the introduction of the relevant noise and jitter parameters and modeling of the jitter probabilities in Section \ref{sec:sysparams_jitterprobs}, we start with the density matrix of the target state, $\ket{\Psi_{\text{T}}}\bra{\Psi_{\text{T}}}$, and model its evolution into $\rho_1$, the state resulting from the target state by incorporating noise and photon loss, as rigorously explained in Section \ref{sec:calcNoiseLossProbs}. Next, this density matrix is used to simulate the click matrices we expect to be obtained for measurements in the Time-of-Arrival- and the Temporal-Superposition-basis (as introduced in Section \ref{sec:setup}) before jitter is accounted for (see Section \ref{sec:calcNoiseLossProbs} and Appendix \ref{app:click_mat_sim}).

Subsequently, a jitter simulation is applied to these click matrices (as described in Section \ref{sec:jitter_simulation}). The resulting jittered click matrices represent the click matrices one would also get from experimental observations.

Those elements of these matrices, which belong to the three theoretical bases, one of which is the basis corresponding to TOA-measurements, $\mathcal{B}_0$, and two of which are the bases corresponding to TSUP-measurements, $\mathcal{B}_1$ and $\mathcal{B}_2$ (see Appendix \ref{app:bases} for details), are then normalized according to their basis belonging, as detailed in Appendix \ref{app:normalization}. Elements of the click matrices that are not directly associated with any of the three bases are discarded. Thus, the `basis-normalized' coincidence click matrices are obtained.

Some of these matrices then are used for the reconstruction of the density matrix $\rho_2$, the density matrix compatible with the (simulated or experimentally observed) jittered click matrices. Note that we actively reconstruct only those elements of the density matrix $\rho_2$ that can be directly inferred from the observed data. These partial reconstructions then serve as constraints for SDPs tailored to lower-bound the key rate, the EoF, or the fidelity. Since these optimize over all remaining unobserved matrix elements, the obtained optimal density matrices may vary depending on the objective quantity, but are of no interest to us, since we solely aim for the lower bounds of the respective performance metrics themselves.

\begin{figure*}[!htbp]
    \centering    \includegraphics[width=0.85\textwidth]{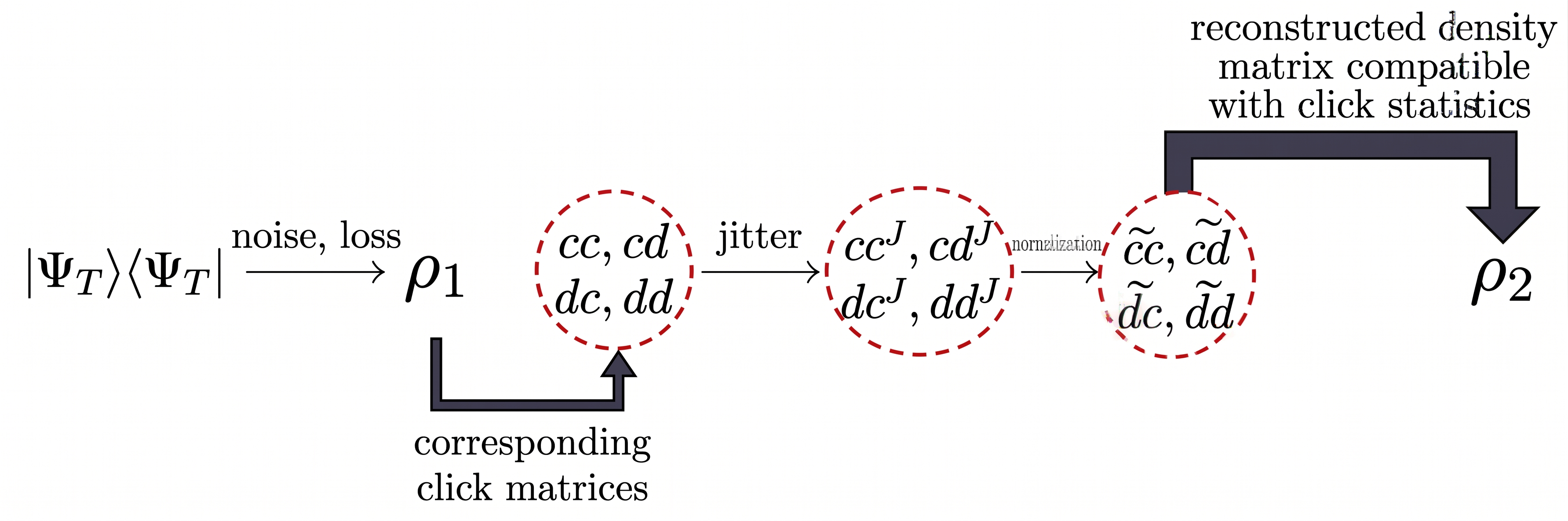}
\caption{Framework of the modeling approach. Starting with the target state $\ketbra{\Psi}{\Psi}$, it is mixed with white noise in a way that already accounts for transmission loss, environmental photons, dark counts and detector inefficiencies. The resulting density matrix $\rho_1$ is used to simulate its corresponding click matrices. These are then manipulated in order to obtain the `jittered' click matrices which account for jitter due to impreciseness of the detectors and the time tagging modules. After normalizing them to comply with the mathematical bases, some state $\rho_2$ representing the `worst-case' density matrix compatible with the simulated click statistics is reconstructed from them. The picture was created with the help of Google Gemini.}\label{fig:modeling_outline}
\end{figure*}

The reconstruction of the diagonal and `first off-diagonal' elements is detailed in Section \ref{sec:reconstruct_density}, while the reconstruction of further off-diagonal-elements for the key rate calculation is explained in Section \ref{subsec:matrix_completion_offdiagonals}, whereas the implicit reconstructions via semi-definite programmes (SDPs) for the EoF and the fidelity are presented in Sections \ref{sec:lower-bounding_EoF} and \ref{sec:lower-bounding_Schmidt} respectively. Section \ref{sec:discussion} discusses the results of the simulations obtained for various parameter configurations. The steps for getting from the target state $\ketbra{\Psi}{\Psi}$ to the density matrix $\rho_2$ are depicted in figure \ref{fig:modeling_outline}. 

\section{Experimental setup and procedure}\label{sec:setup}
While the presented noise and jitter model can be applied to a broad range of experimental setups, its applications requires a rigorous analysis of the underlying setup and the reconstruction of the density matrix of the state shared by Alice and Bob. For the construction of the click matrices and their normalization, we build upon the framework established in \cite{BergKan}, outlining only the main points and refinements here. Details on some exemplary experimental implementation can be found in \cite{Schiffer_2026}.
\begin{figure*}[!htbp]
    \centering
    \includegraphics[width=1.\textwidth]{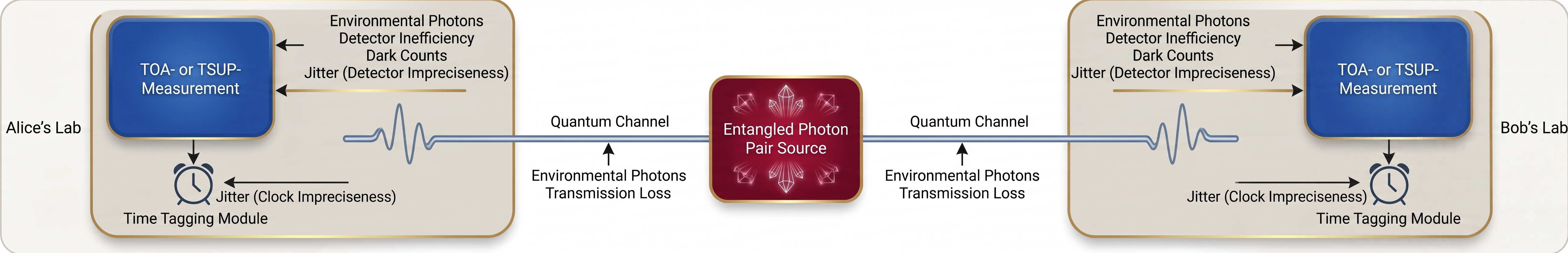}
\caption{Scheme of the setup considered and the sources of noise affecting it. Entangled photon pairs are emitted by the central source and transmitted via a quantum channel, where they are subject to transmission loss and background noise from environmental photons. Within the parties' laboratories, additional environmental photons can enter the setup. The detectors introduce further noise via dark counts, can fail to register photons due to finite detector efficiency, or can assign detection events to the wrong time-bin due to detector jitter. Finally, photons may also be assigned to an incorrect time-bin due to clock imprecision in the time-tagging modules, which is likewise accounted for as timing jitter.. The picture was created with the help of Google Gemini.}
\label{fig:sketch_setup}
\end{figure*}

\begin{figure*}[!htbp]
    \centering
    \includegraphics[width=1.\textwidth]{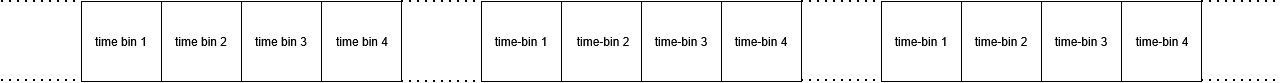}
\caption{Example for dimension four: Each time-frame is built out of four time-bins. The time-frames are separated by inter-frame pauses in the size of one time-bin.}
\label{fig:frame_sketch}
\end{figure*}

Two identical measurement devices are placed in the laboratories of the two parties. An entangled photon pair source, which does not need to be trusted, is positioned either in between the two labs or in one of them. All our considerations apply in any case. (For a sketch of the setup and the sources of noise affecting it, see Figure \ref{fig:sketch_setup}. For a scheme depicting the physical laboratory objects involved, see Figure 1 in \cite{BergKan}.) The parties' time-reference points are defined via synchronized coincidence windows, so-called `time-bins' which are grouped into `time-frames', see Figure \ref{fig:frame_sketch}. 
Any offset in arrival times due to asymmetric distances from the source to the respective detectors is compensated in the setup or data post-processing to ensure temporal agreement.

After discretizing the temporal degree of freedom into $d \in \mathbb{N}_{\geq 2}$ time-bins, the polarized, time-entangled photon pair produced by the source reads:
\begin{equation}    \ket{\Psi}:=\sum\limits_{p_A, p_B \in \{H,V\}} c_{p_A, p_B} \ket{p_A, p_B} \otimes \frac{1}{\sqrt{d}} \sum\limits_{k=0}^{d-1}\ket{kk},
\end{equation}
where $c_{p_A,p_B} \in \mathbb{C}$ satisfies $\sum\limits_{p_A, p_B} |c_{p_A,p_B}|^2 = 1$ and $H$ and $V$ denote the two possible polarizations orthogonal to each other, `horizontal' and `vertical', while $\ket{kk}$ represents the joint state where both photons occupy the $k$-th time-bin in the TOA basis. The shares of the photon pair are transmitted to Alice and Bob respectively.

Each lab is equipped with a measurement
setup where incoming photons pass
a $\theta:1-\theta$ beam splitter, which, with probability $\theta$, directs them to the Time-of-Arrival (TOA) arm to allow
for measurements in the TOA basis, capturing the precise arrival times, and, with probability $1-\theta$, leads them into the Temporal-Superposition (TSUP) arm, which contains a polarizing imbalanced interferometer to execute the superposition of two neighboring time-bins, enabling measurements in the TSUP basis. (Note that the setup analysis can be extended also to other interferometer lengths, as done e.g. in \cite{Schiffer_2026}.) While we assume a passive basis choice as a robust simplified baseline for security analysis, the model and analysis remain valid for active basis choice, provided the intended $\theta:1-\theta$ probability distribution is maintained. 

Now, we want to modify the shared state to account for photon loss on their way from the source to the detectors, environmental noise such as background photons from sunlight, dark counts (i.e. detector clicks in the absence of any photon) and detector inefficiencies causing missed clicks. To this end, we focus exclusively on the temporal part of the target state, reading 
\begin{equation}\label{equ:target_state_time}
    \ket{\Psi_{T}}:=\frac{1}{\sqrt{d}}\sum\limits_{k=0}^{d-1}\ket{kk}.
\end{equation} 
(This focus is justified by the fact that polarization and time-of-arrival can be treated as independent degrees of freedom. As explained in \cite{BergKan}, by locally projecting the polarization onto a fixed state (e.g. $\ket{D}$) prior to detection, a tensor product structure of the form $\rho_{\text{pol}}\otimes \rho_{\text{time}}$ is ensured, allowing the polarization, which only serves as an ancilla for interferometry and doesn't contribute to the information Alice and Bob exchange, to be traced out and excluded in our subsequent analysis.)

The measurements performed by Alice and Bob over the experimental procedure result in clicks with time-stamps which are stored in four kinds of so-called `coincidence click matrices', denoted \begin{equation}
    cc(i,j), \, dd_{a,b}(i,j), \, cd_{b}(i,j), \, dc_{a}(i,j),
\end{equation} 
where $a,b \in \{1,2\}$ label the detectors and $i,j \in \{0, \dotsc, d-1\}$ the time stamps. The notations $c$ and $d$ are derived from `computational' (TOA) and `diagonal' (TSUP), respectively. Note that by construction of the coincidence click matrices, the data is post-selected on coincidence clicks, and both single click and multi-click events are discarded.

Details on the click matrices are given in Appendix \ref{app:click_mat_sim} and details on the mathematical representation of the measurement operators involved in the setup can be found in \cite{BergKan}.

\section{System parameters and jitter probabilities}\label{sec:sysparams_jitterprobs}
We now introduce the parameters considered in our model, which consist of the `tunable parameters' which can be adjusted in experiments and are listed in Section \ref{subsec:tunableParams}, further setup parameters which depend on the devices used and environmental conditions, including also jitter caused by detector and clock impreciseness, listed in Section \ref{subsec:noise_jitter}. A summary of all our model parameters is given in Table \ref{tab:parameters}. In Section \ref{subsec:jitter_probs_calc} we finally explain how to model the jitter probabilities corresponding to jitter given in units of time.

\subsection{Tunable parameters}\label{subsec:tunableParams}
Here we give an overview of the parameters that can be adjusted in experiments and which can therefore be optimized a priori using our noise model. 
\begin{itemize}
    \item Dimension $d$: The dimension of the temporal Hilbert space of both Alice and Bob. It represents the number of time-bins within a single time-frame (e.g. $d = 4$ corresponds to the local Hilbert space $\mathbb{C}^4$ for each party, resulting in a joint state described by a $16 \times 16$ density matrix acting on the composite Hilber space $\mathbb{C}^4 \otimes \mathbb{C}^4$).
    \item Bin size $\tau$: The temporal duration of a single time-bin in seconds (e.g. $\tau = 10^{-9}$ s). The multiplication of the bin size with the dimension yields the Frame size $\text{T}$, defined by $T:=d \cdot \tau$, giving the total duration of a single time-frame in seconds. In other words, a frame represents the temporal window spanned by the full sequence of $d$ orthogonal time-bins defined in the TOA basis. Note that $T$ refers strictly to the information-encoding interval. Pauses for avoiding inter-frame crosstalk are modeled external to the frame structure.
    \item Length of inter-frame pauses $\text{l}_{\text{p}}$: To prevent crosstalk between neighboring frames, they are separated by guard intervals during which no clicks are recorded. In our simulations, a pause duration in the source, equal to a single time-bin is used, as this duration turns out to be sufficient, since the probability of a photon jittering by more than one time-bin length is negligible under our configurations.\footnote{One could also just remove the clicks in the guard window period in post-processing, but then additional jitter events would be recorded at the boundary. For our results it doesn't make a qualitative difference.} 
    \item Pair production rate $\gamma$: Let $P_{\text{prod}}(n(t))$ be the probability that exactly $n$ photon pairs - hereafter referred to as `source photon pairs' (or individually as `source photons') - are emitted by the source within a time interval of $t$ seconds.
    
   As the photon pairs are generated independently and at a constant average rate, we assume an underlying Poisson process $t\mapsto n(t)$ where the expectation value for the number of photon pairs produced per second is given by the pair production rate $\gamma$ (e.g. $\gamma = 10^7$ $s^{-1}$) and the expectation value for the number of photon pairs produced per time-frame is given by $\gamma\cdot\text{T}$.
   
    Formally, this means $P_{\text{prod}}(n(t)) := P_{\gamma\cdot \text{t}}(n)$ (with $P_{\lambda}$ denoting the Poisson distribution with Poisson parameter $\lambda$, such that $P_{\lambda}(n):=\frac{(\lambda)^n e^{-\lambda}}{n!}$).
    \item TOA-TOA coincidence probability $p_{\text{CC}}$: As already mentioned, we assume a passive basis choice via a $\theta:1-\theta$ beam splitter (or an active basis choice with equivalent probability distribution), directing the photon to a TOA-measurement with probability $\theta$. Hence the probability for a TOA-TOA-measurement is $p_{\text{CC}}=\theta^2$.
    
    While $p_{\text{CC}}$ doesn't affect the key rate per coincidence click, it massively impacts the key rate per second, since only TOA-TOA coincidences contribute to the key. However, $p_{\text{CC}}$ cannot be chosen arbitrarily high, because for security reasons a sufficient amount of TSUP clicks is required (as detailed in \cite{KanDual2}).
\end{itemize}

\subsection{Noise and jitter}
\label{subsec:noise_jitter}
Unlike the experimental settings described previously, the following noise parameters are governed by physical and environmental limitations. They can only be influenced by significant changes in the experimental setup or further technological developments, for example by using superior detectors, optimizing the source location, or improving the shielding against environmental photons.

\begin{itemize}
    \item Detector efficiency $\eta_A, \eta_B$: The probability that a detector on Alice's resp. Bob's side registers a click when a photon is present.
    \item Transmission loss probability $P_{\text{loss}}^A, P_{\text{loss}}^B$: The probability of a photon being lost during transmission to the respective lab (e.g. if the source is located in Alice’s lab, $P_{\text{loss}}^A\approx 0$, whereas a central location might yield $P_{\text{loss}}^A = P_{\text{loss}}^B = 0.997$). Consequently, $P_{\text{surv}}^A:=1-P_{\text{loss}}^A$ and $P_{\text{surv}}^B:=1-P_{\text{loss}}^B$ are the probabilities for a photon to survive transmission through Alice's resp. Bob's channel.
    
    \item Environmental photon rate $\kappa_A, \kappa_B$: We account for the number of environmental photons we expect to reach the detector per second, where we assume this number to follow a Poisson process, which means that the probability that exactly $n$ environmental photons are present in Alice's lab within a time-frame of length $\text{T}$ is given by $P_{\text{env}}^A(n):=P_{\kappa_A \cdot \text{T}}(n)$ (where $P_{\lambda}$ denotes the Poisson distribution with Poisson parameter $\lambda$). 
    \item Dark count rate $\zeta_A, \zeta_B$:  We account for the number of detector clicks we expect per second in the absence of any photon, where we assume this number to follow a Poisson process. Similar to the environmental photon rate, the probability that Alice records exactly $n$ dark counts within a time-frame of length $\text{T}$ is given by $P_{\text{dark}}^A(n):=P_{\zeta_A\cdot \text{T}}(n)$. 
\end{itemize}
The jitter - which results in the incorrect registration of a photon in a neighboring time-bin instead of the time-bin in which it was intended to be detected - is a critical parameter which cannot be chosen arbitrarily either, but is inherently linked to the bin size. Smaller time-bins naturally allow higher dimensions within the same time-frame, which would go in hand with higher key rates. But as due to jitter, the probability for a photon to be registered in a neighboring bin (due to clock impreciseness or delayed clicking a detector) increases with decreasing bin size, there's a fundamental trade-off between these two parameters: Smaller bin sizes provide higher potential dimensionality but suffer from increased jitter and noise, both of which degrade the key rate.

A central achievement of our model is the ability to navigate this trade-off by finding a highly efficient configuration with regard to experimental optimization in advance.

We consider the following jitter probabilities for each party $j \in \{A,B\}$:
\begin{itemize}
    \item $p_{jd}^J$, the probability that a detected photon is registered in the bin following its correct arrival bin.
    \item $p_{je}^J$, the probability that a photon is registered in the bin preceding the correct one.
\end{itemize}

\begin{table*}[htbp]
\centering
\label{tab:parameters}
\begin{tabular}{lll}
\hline\hline
\textbf{Symbol} & \textbf{Parameter description} & \textbf{Unit; Details} \\ \hline
$d$ & dimension of the local temporal Hilbert space & dimensionless \\
$\tau$ & bin size & seconds \\
$T$ & frame size & seconds; $T = d \cdot \tau$ \\
$\text{l}_{\text{p}}$ & length of inter-frame pauses for crosstalk prevention & seconds; set to the length of one time-bin \\
$\gamma$ & pair production rate of the source & number of pairs per second \\
$P_{\text{prod}}(n(t))$ & probability of generating exactly $n$ photon pairs during $t$ seconds & dimensionless; $0 \leq P_{\text{prod}}(n(t)) \leq 1$ \\
$\theta$ & probability of directing a local photon to a TOA measurement & dimensionless; $0 \leq \theta \leq 1$; set to $\theta = 0.75$ \\
$p_{\text{CC}}$ & probability of a joint TOA-TOA-measurement & dimensionless; $p_{\text{CC}} = \theta^2$ \\
$\eta_A/\eta_B$ & detector efficiency on Alice's/Bob's side & dimensionless; $0 \leq \eta_A, \eta_B \leq 1$\\
$P_{\text{loss}}^A/P_{\text{loss}}^B$ & loss probability on Alice's/Bob's side & dimensionless; $0 \leq P_{\text{loss}}^A, P_{\text{loss}}^B \leq 1$\\
$P_{\text{surv}}^A/P_{\text{surv}}^B$ & probability for a source photon of surviving the transmission to Alice/Bob & dimensionless; $P_{\text{surv}}^j = 1 - P_{\text{loss}}^j, \, j \in \{A,B\}$\\
$\kappa_A/\kappa_B$ & environmental photon rate on Alice's/Bob's side & number of photons per second \\
$\zeta_A/\zeta_B$ & dark count rate on Alice's/Bob's side & number of dark counts per second \\
$p_{Ad}^J/p_{Bd}^J$ & probability for a click on Alice's/Bob's side to be delayed due to jitter & dimensionless; $0 \leq p_{Ad}^J, p_{Bd}^J \leq 1$ \\ $p_{Ae}^J/p_{Be}^J$ & probability for a click on Alice's/Bob's side to be early due to jitter & dimensionless; $0 \leq p_{Ae}^J, p_{Be}^J \leq 1$ \\
$n_{\text{frames}}$ & frame rate & number of frames per second\\
\hline\hline
\end{tabular}
\caption{Overview of the main parameters involved in the experimental setup and their physical meaning.}
\end{table*}

\subsection{Modeling Jitter}
\label{subsec:jitter_probs_calc}
In experimental contexts, jitter is typically characterized in units of time (e.g. seconds or picoseconds) rather than probabilities. A very simplified approach to model jitter is to assume that the timing of the detection of a photon that reaches the detector within a certain time-bin follows a Gaussian distribution $\mathcal{N}(\mu, \sigma^2)$ with the mean $\mu$ being centered in the middle of the time-bin and the standard deviation $\sigma$ being referred to as jitter.

Then, the conversion from jitter in units of time to the probabilities $p_{Ae}^J, p_{Ad}^J, p_{Be}^J, p_{Bd}t^J$ is done by setting $p_{Ae}^J = p_{Ad}^J$ resp. $p_{Be}^J = p_{Bd}^J$ (because of the mean being assumed to be in the center of the time-bin) and calculating these probabilities under usage of the cumulative distribution function (cdf) of the Gaussian distribution $\mathcal{N}(\frac{\tau}{2}, {\sigma_A}^2)$ resp. $\mathcal{N}(\frac{\tau}{2}, {\sigma_B}^2)$.\\
For example, given a jitter-$\sigma_A$ of $10^{-10}$ seconds, we have the jitter probability 
\begin{equation*}
p_{Ae}^J = p_{Ad}^J = \frac{1-(\text{cdf}(\tau, \frac{\tau}{2}, \sigma_A) - \text{cdf}(0, \frac{\tau}{2}, \sigma_A))}{2},
\end{equation*} where $\text{cdf}(x, \mu, \sigma)$ is the cumulative distribution function corresponding to $\mathcal{N}(\mu, \sigma^2)$, evaluated at point $x$.

However, this assumption of perfect centering is often unjustifiable and could lead to an overestimation of the achievable entanglement and key rate. That simplification is only legitimate if one works with extremely small or very large time-bins, but for our model especially the intermediate cases are of interest, as they could correspond to the optimal regime.

Therefore, here we take a more conservative approach: We treat $\mu$ as a variable and average over associated Gaussian distributions. More precisely, we suggest the following procedure: 
\begin{enumerate}
    \item
        \begin{enumerate}
        \item Determine the maximal interval $[y_{min}, y_{max}]$ such that $\text{cdf}(\tau,y,\sigma_A) - \text{cdf}(0,y,\sigma_A) \geq p_{\epsilon}$ for $y \in [y_{min}, y_{max}]$, where $p_{\epsilon}$ is a fixed value in $(0,1)$. 
        This means, one finds all the values for the mean $y$ of the distribution $\mathcal{N}(y,\sigma_A^2)$, for which the probability that the click is detected within the interval $[0, \tau]$ is `significant', namely at least $p_{\epsilon}$. (Be aware that it does not make sense to consider mean values $y$ for which the probability for the click to be detected inside $[0,\tau]$ is arbitrarily small (i.e. much smaller than $0.5$, if not even nearby $0$), because then in the end one would deal with the average over Gaussian distributions with mean values throughout $\mathbb{R}$, which would mean that regardless of the value of $\sigma_A$ the limits $p_{Ae}^J = p_{Ad}^J \approx 0.5$ would be obtained, so the probability for jitter would be around $1$, a proposition which would be of no use.)
        
        The Gaussian distributions $\mathcal{N}(y_{min},\sigma_A^2)$ and $\mathcal{N}(y_{max}, \sigma_A^2)$ are then the `worst-case-distributions'  (in the sense that all $\mathcal{N}(\mu, \sigma_A^2)$ with $y_{min} < \mu < y_{max}$ have a higher probability for their corresponding detection time to lie within $[0, \tau]$) that we take into account.
        \item Alternatively, another sensible approach would be to not determine the interval for $y$ such that $\text{cdf}(\tau,y,\sigma) - \text{cdf}(0,y,\sigma) \geq p_{\epsilon}$, but to consider only the Gaussian distributions the mean of which lies within $[0,\tau]$, which means that the interval of interest would be $[y_{min} = 0, y_{max} = \tau]$.
        \end{enumerate}
        
        Since both approaches are legitimate, one could use a combination of both in the sense that one calculates the jitter probabilities with both of them and then uses the worse (i.e. larger) of both jitter probabilities $p_{Ae}^J + p_{Ad}^J$ for further calculations.
        
        \item[] However, if we set $p_{\epsilon} = 0.5$ in the first method, which is a natural choice because then for all the Gaussian distributions associated with it, it is more likely that a click is detected within the interval $[0, \tau]$ than outside of it, then (as long as $[y_{min}, y_{max}]$ is not empty, which can happen if the jitter-$\sigma$ is large compared to the bin size), we have $y_{min} = 0$ and $y_{max} = \tau$ (apart from computational impreciseness), so that both approaches coincide. 
        
        \item[] Therefore, in our calculations we are using only the second method, presented in 1. (b) as it corresponds to the choice $p_{\epsilon} = 0.5$ for the first one, but significantly accelerates the simulation.
    \item Integrate $\text{cdf}(0, y, \sigma_A)$ and $1-\text{cdf}(\tau,y,\sigma_A)$ over the interval $[y_{min},y_{max}]$ and divide both results by the interval length $y_{max} - y_{min}$, which yields \begin{equation*}p_{Ae}^J := \frac{1}{y_{max}-y_{min}}\int\limits_{y_{min}}^{y_{max}} \text{cdf}(0,y,\sigma_A)dy\end{equation*} and \begin{equation*}p_{Ad}^J :=  \frac{1}{y_{max}-y_{min}}\int\limits_{y_{min}}^{y_{max}} 1-\text{cdf}(\tau,y,\sigma_A)dy.\end{equation*}
    \item[] Note that due to the symmetry of the Gaussian distribution, also with these approaches we expect that $p_{Ad}^J = p_{Ae}^J$ in theory, but for computational limitations (like floating-point-precision, numerical integration, determination of $[y_{min}, y_{max}]$ via discretization, ...) it can happen that they are not the exact same value.
    \item Repeat steps $1$ and $2$ using $\sigma_B$ instead of $\sigma_A$ to obtain $p_{Be}^J$ and $p_{Bd}^J$.
    \item [] By avoiding the naive `center-bin' assumption, we obtain a jitter estimation that is significantly larger and more realistic for practical experimental parameters. This underlines how important it is to model the jitter in a sensible way without making non-justifiable assumptions, ultimately yielding a secure lower bound on the key rate and entanglement rate. 
\end{enumerate}

\section{Theoretical foundations of the key rate and quantification of entanglement}
In this section, we introduce our performance quantities of interest, which are calculated in our simulations: The key rate in Section \ref{subsec:key_theory}, followed by the Entanglement of Formation and the related entanglement rate in Section \ref{subsec:EoF_theory} and the fidelity and Schmidt number in Section \ref{subsec:Schmidt_theory}.
\subsection{Key rate}\label{subsec:key_theory}
The primary figure of merit for QKD applications is the asymptotic secret key rate, which quantifies the number of secure bits that can be extracted per round of entanglement distribution.

in the asymptotic limit of infinitely many rounds and under the assumption of collective attacks, it has been established in \cite{Devetak_Winter} that the asymptotic key rate can be lower-bounded via 
\begin{equation*}
    r\geq I(A:B) - I(A:E),
\end{equation*}
where $I(A:B)$ represents the mutual information shared by Alice and Bob and $I(A:E)$ denotes the information accessible to Eve about Alice’s part of the raw key (see Ref. \cite{Wolf_2021}).

In our analysis, we focus on the key rate per coincidence click, $r_{cc}$. This is obtained by replacing the signal-based information quantities with their coincidence-based counterparts, $I(A:B)_{cc}$ and $I(A:E)_{cc}$. These terms describe the mutual information shared or leaked specifically when both parties register a coincidence click.

However, for high-dimensional QKD, the key rate per coincidence click, which is naturally bounded by $\text{log}_2(d)$, is not the most suitable metric for comparing the performance of different dimensions. While higher dimensions theoretically allow for more information per click, they may also lower coincidence rates, since only time-frames with exactly one coincidence click contribute to the key while frames with none or multi-click-events have to be discarded. Therefore, we focus on the key rate per second, $r_{\text{sec}}$, as our quantity of interest. It tells us the amount of secure key that Alice and Bob obtain per second and it can easily be converted from and into the key rate per coincidence click, as explained in Appendix \ref{app:keyrate_per_second}.

\subsection{Entanglement of Formation and Entanglement rate}\label{subsec:EoF_theory}
Beyond the key rate, we aim to quantify the entanglement of the generated time-bin entangled state. This in general is a non-trivial task, especially since we cannot reconstruct a full density matrix $\rho$, but only its main diagonal and (the real part of) certain off-diagonal elements, as we will see in more detail in Section \ref{sec:reconstruct_density}.

Our primary metric is the Entanglement of Formation (EoF), an entanglement measure which allows an operationally meaningful interpretation, as it quantifies the number of maximally entangled qubit pairs, called `ebits', required to prepare $\rho$ using arbitrary local operations and classical communication (LOCC)~\cite{Bennett_1996, Horodecki_2009}.

As an entanglement measure, the EoF of any $d$-dimensional state $\rho$ lies between $0$ and $\text{log}_2(d)$. A practical lower bound for the EoF is given by \cite{Huber_2013, Martin_2017} \begin{equation}\label{equ:EoF_bound}
    \text{EoF}(\rho) \geq -\text{log}_2(1-\frac{B(\rho)^2}{2}),
\end{equation}
where $B(\rho)$ is defined as 
\begin{equation}\label{equ:C_in_EoF_bound}
    B(\rho) = \frac{2}{\sqrt{|C|}}\sum\limits_{\substack{(j,k)\in C \\ j < k}}\left( |\bra{jj|\rho}\ket{kk}| - \sqrt{\bra{jk|\rho}\ket{jk}\bra{kj|\rho}\ket{kj}}\right),
\end{equation}
in which $C$ is a subset of the set of all index pairs that can be chosen arbitrarily and $|C|$ denotes its cardinality. To obtain the tightest possible lower bound, $C$ must be chosen to maximize $B(\rho)$.

However, the number of pairs that can be included in $C$ is limited by our knowledge of the off-diagonal elements of $\rho$. To address this, we employ an SDP tailor-made for lower-bounding the EoF, which takes the (known) diagonal and the (lower-bounded) off-diagonal of the density matrix as input and minimizes the EoF over all density matrices compatible with these constraints (see Section \ref{sec:lower-bounding_EoF}).

Note that due to the squaring in equation \eqref{equ:EoF_bound}, it is important to keep in mind that negative values for $B(\rho)$ do not make sense and would yield wrong lower bounds on the EoF due to the squaring in equation \eqref{equ:EoF_bound}. This becomes obvious when interpreting $B(\rho)$ as the lower bound on the concurrence of $\rho$ \cite{Martin_2017}, which is an entanglement monotone and therefore lies in the range $[0, \text{log}_2(d)]$, for any $d$-dimensional state $\rho$ \cite{Friis_Book}. Therefore, whenever the maximization of $B(\rho)$ yields a negative value, $B(\rho)$ has to be set to zero.

As the EoF measures entanglement per coincidence click, the so-called entanglement rate, which can be interpreted as the number of ebits exchanged per second (instead of per coincidence click), is more suited for the comparison of entanglement in the course of the protocol over various dimensions. The entanglement rate, denoted $\text{EoF}_{sec}$ is obtained from the entanglement of formation, denoted $\text{EoF}_{cc}$, by multiplying it with the TOA-TOA-coincidence click rate (per second), meaning \begin{equation*}
    \text{EoF}_{\text{sec}} = \text{EoF}_{\text{cc}} P_{\text{click}} p_{\text{CC}}\, n_{\text{frames}},
\end{equation*} where $P_{\text{click}}$ is the probability that exactly one coincidence click occurs within a frame, while $p_{\text{CC}}$ is the probability for a measurement to take place in the TOA-TOA-basis and $n_{\text{frames}}$ is the frame rate, i.e. the number of frames per second. The derivation of the formula is detailed in Appendix \ref{app:keyrate_per_second}.

\subsection{Fidelity and Schmidt number}\label{subsec:Schmidt_theory}
In addition to the key rate and EoF, we determine a lower bound for the fidelity $\mathcal{F}(\rho, \ket{\Psi})$, defined as 
\begin{equation}\label{equ:fidelity_definition}
\mathcal{F}(\rho, \ket{\Psi}):=\braket{\Psi|\rho}{\Psi}.
\end{equation} 
for a pure state $\ket{\Psi}$. The fidelity naturally ranges from $0$ to $1$, where a higher value indicates a greater similarity between two states. Physically, $\mathcal{F}(\rho, \ket{\Psi})$ expresses the probability that a projective measurement of $\rho$ results in $\ket{\Psi}$. Moreover, it serves as an entanglement witness for estimating the Schmidt number.

Our objective is to evaluate $\mathcal{F}(\rho_2, \ket{\Psi}_T)$, the fidelity of the reconstructed state compatible with the simulated measurements to the maximally entangled target state defined in Equ. \eqref{equ:target_state_time}, $\ket{\Psi_T}$.

For this application, we utilize the property (from \cite{Terhal_2000}, Lemma 1) that for any density matrix $\rho$ on $\mathcal{H}_d \otimes \mathcal{H}_d$ with Schmidt number $k$, the so-called maximal fidelity $\mathcal{F}_{\text{max}}(\rho)$ fulfills 
\begin{equation}\label{equ:max_fidelity_definition}
    \mathcal{F}_{\text{max}}(\rho):=\underset{\Phi}{\text{max}} \braket{\Phi|\rho}{\Phi} \leq \frac{k}{d},
\end{equation}
where the maximization runs over all maximally entangled, pure states $\ket{\Phi}$. (\cite{Friis_Book})

Consequently, if the fidelity observed for a density matrix $\rho$ and some maximally entangled, pure state $\ket{\Psi}$ satisfies $\braket{\Psi|\rho}{\Psi} > \frac{k}{d}$, then $\rho$ must have a Schmidt number of at least $k+1$.

\section{Coincidence click probabilities and the resulting density matrix} \label{sec:calcNoiseLossProbs}
To integrate the parameters described in Section \ref{sec:sysparams_jitterprobs} into our model, we must account for their influence on the emergence of a coincidence click, which is defined as exactly one detection event for each party within the same time-frame. Therefore, in this section, we determine the probabilities for a coincidence click to originate from each possible physical event. We end this section with the calculation of the state expected to reach the detectors prior to the influence of jitter.

These possible events yielding a coincidence click are categorized as follows: 
\begin{itemize}
    \item Correlated source pair: The coincidence is caused by a single entangled photon pair. Note that this is the only `good' case. we denote this probability by $p(A\varphi, B\varphi)$. 
    The detected state associated with this probability is $\ketbra{\psi_T}, $ where $\ket{\psi_T} = \frac{1}{\sqrt{d}}\sum\limits_{j=0}^{d-1} \ket{jj}$.
    
    \item Uncorrelated source photons: The coincidence stems from a photon of one source pair and a second photon from a distinct source pair. This probability is denoted by $p(A\varphi, B\tilde{\varphi})$.
    The associated state reads $\text{Tr}_B\ketbra{\psi_T} \otimes \text{Tr}_A \ketbra{\psi_T} = \frac{1}{d}\mathbb{1}_{d\times d}\otimes \frac{1}{d}\mathbb{1}_{d\times d} = \frac{1}{d^2}\mathbb{1}_{d^2\times d^2}$. 
    
    \item Source and environmental photons: The coincidence is caused by one source photon and one environmental photon. The corresponding probabilities are denoted by $p(A\varphi, B\xi)$ and $p(A\xi, B\varphi)$, depending on which party detects the source photon. The states associated with these cases are $\text{Tr}_B\ketbra{\psi_T}\otimes \frac{1}{d}\mathbb{1}_{d\times d} = \frac{1}{d^2}\mathbb{1}_{d^2\times d^2}$ and $\frac{1}{d}\mathbb{1}_{d\times d} \otimes \text{Tr}_A\ketbra{\psi_T}= \frac{1}{d^2}\mathbb{1}_{d^2\times d^2}$.
  
    \item Source photon and dark count: The coincidence stems from a source photon and a detector dark count, the probabilities being denoted by $p(A\varphi, B\delta)$ and $p(A\delta, B\varphi)$, depending on which party detects the source photon. The corresponding states read $\text{Tr}_B\ketbra{\psi_T}\otimes \frac{1}{d}\mathbb{1}_{d\times d} = \frac{1}{d^2}\mathbb{1}_{d^2\times d^2}$ and $\frac{1}{d}\mathbb{1}_{d\times d} \otimes \text{Tr}_A\ketbra{\psi_T} = \frac{1}{d^2}\mathbb{1}_{d^2\times d^2}$.
    
    \item Environmental photon and dark count: The coincidence click is caused by an environmental photon and a dark count. The associated probabilities are denoted $p(A\xi, B\delta)$ and $p(A\delta, B\xi)$, depending on which party detects a dark count. The corresponding state reads $\frac{1}{d}\mathbb{1}_{d \times d} \otimes \frac{1}{d}\mathbb{1}_{d \times d} = \frac{1}{d^2}\mathbb{1}_{d^2 \times d^2}$.

    \item Two dark counts: The coinciding clicks arise from detector dark counts. The corresponding probability is denoted $p(A\delta, B\delta)$ and the associated state is given by $\frac{1}{d}\mathbb{1}_{d \times d} \otimes \frac{1}{d}\mathbb{1}_{d \times d} = \frac{1}{d^2}\mathbb{1}_{d^2 \times d^2}$.
    
    \item Two environmental photons: Both detections originate from environmental photons, the corresponding probability being denoted $p(A\xi, B\xi)$. The associated state is $\frac{1}{d}\mathbb{1}_{d \times d} \otimes \frac{1}{d}\mathbb{1}_{d \times d} = \frac{1}{d^2}\mathbb{1}_{d^2 \times d^2}$.
\end{itemize}

Notably, except for the `good' case of correlated source photons, all events result in a state characterized by so-called white noise, which is represented by the maximally mixed state $\frac{1}{d^2}\mathbb{1}_{d^2 \times d^2}$. This is in accordance with the fact that all sources of noise relevant in a free space experiment stem from random processes which are best modeled in this manner.

By analyzing these possibilities using basic combinatorial methods, we derive the explicit formulas for the probabilities, as presented in Appendix \ref{app:prob_calc}. Summing the products of these probabilities and their associated density matrices yields the state expected to reach the detectors prior to the influence of jitter, $\rho_1$, via: 
\begin{equation}
    \rho_1:=\frac{1}{\text{tr}(\rho'_1)}\rho'_1,
\end{equation}
where
\begin{widetext}
\begin{equation}\label{equ:rho1'}
    \begin{aligned}
        \rho'_1:&=p(A\varphi, B\varphi)\frac{1}{d}\sum\limits_{k=0}^{d-1}\sum\limits_{j=0}^{d-1}\ket{jj}\bra{kk} + p(A\varphi, B\tilde{\varphi})\frac{1}{d^2}\mathbb{1}_{d^2\times d^2} + (p(A\varphi, B\xi) + p(A\xi, B\varphi))\frac{1}{d^2}\mathbb{1}_{d^2 \times d^2} \\&+ (p(A\varphi, B\delta)+p(A\delta, B\varphi)\frac{1}{d^2}\mathbb{1}_{d^2\times d^2} + (p(A\xi, B\delta) + p(A\delta, B\xi))\frac{1}{d^2}\mathbb{1}_{d^2 \times d^2} + p(A\delta, B\delta)\frac{1}{d^2}\mathbb{1}_{d^2 \times d^2} + p(A\xi, B\xi)\frac{1}{d^2}\mathbb{1}_{d^2 \times d^2}\\
        &= p(A\varphi, B\varphi)\frac{1}{d}\sum\limits_{k=0}^{d-1}\sum\limits_{j=0}^{d-1}\ketbra{jj}{kk} + (p(A\varphi, B\tilde{\varphi}) + p(A\varphi, B\xi) + p(A\xi, B\varphi) + p(A\varphi, B\delta) + p(A\delta, B\varphi) \\&+ p(A\xi, B\delta) + p(A\delta, B\xi) + p(A\delta, B\delta) + p(A\xi, B\xi))\frac{1}{d^2}\mathbb{1}_{d^2 \times d^2}
    \end{aligned}
\end{equation}
\end{widetext}
is a sub-normalized density matrix. 

As a subsequent step, we have to account for temporal jitter during detection, which may shift the arrival times for one party or both parties.

\section{Simulating jitter}\label{sec:jitter_simulation}
In this section, we present the jitter simulation by constructing so-called `jittered click matrices'. We conclude by explaining how to transform the obtained matrices into valid `basis-normalized' jittered click matrices which can be used for the reconstruction of the density matrix.
\subsection{Simulating the jittered click matrices}
With the aim of obtaining the click matrices that account for jitter, which represent those we would expect to obtain experimentally, we first make use of $\rho_1$ to simulate the unjittered click matrices. For that, we generate the click matrices $cc, dd_{i,j}, cd_i$ and $dc_i$, for $i,j \in \{1,2\}$, introduced in Section \ref{sec:setup} and defined in Appendix \ref{app:click_mat_sim}, which contain the probabilities for coincidence clicks in the time-bins associated with the rows and columns when measuring in the respective bases.

The jitter mechanism is modeled by considering how temporal shifts redistribute these probabilities. For instance, $cc(m,n)$ represents the event where Alice and Bob would have registered a non-jittered coincidence in bins $m$ and $n$ respectively, $m,n \in \{1, \dotsc, d\}$. If Alice’s detection is delayed - an event we assume to be sufficiently unlikely to exceed one bin duration - the click instead contributes to $cc(m+1,n)$, where for $m+1 > d$ the coincidence click is considered lost.

Consequently, the entries of the jittered matrices are calculated by weighting the entries of the unjittered matrices with the probabilities of various jitter events. For the entries of the jittered $cc-$matrix this yields 
\begin{equation}
    \begin{aligned}
        cc^J&(m,n) \\
        &:= (1-p_{Ad}^J-p_{Ae}^J)(1-p_{Bd}^J-p_{Be}^J)cc(m,n)\\ 
        &+ p_{Ae}^J(1-p_{Bd}^J-p_{Be}^J)cc(m+1,n)\\ 
        &+ p_{Ad}^J(1-p_{Bd}^J-p_{Be}^J)cc(m-1,n)\\
        &+ (1-p_{Ad}^J-p_{Ae}^J)p_{Be}^Jcc(m,n+1)\\ 
        &+ (1-p_{Ad}^J-p_{Ae}^J)p_{Bd}^Jcc(m,n-1)\\ 
        &+ p_{Ae}^Jp_{Be}^Jcc(m+1,n+1)\\ 
        &+ p_{Ad}^Jp_{Bd}^Jcc(m-1,n-1)\\ 
        &+ p_{Ae}^Jp_{Bd}^Jcc(m+1,n-1)\\ 
        &+ p_{Ad}^Jp_{Be}^Jcc(m-1,n+1),\, m,n = 2, \dotsc, d-1, 
    \end{aligned}
\end{equation}
where $p_{Ad}^J, p_{Ae}^J, p_{Bd}^J, p_{Be}^J$ are the jitter probabilities defined in Section \ref{subsec:noise_jitter} and calculated as explained in Section \ref{subsec:jitter_probs_calc}. The jittered matrices $dd^J_{ab}, cd^J_{b}, dc^J_{a} \, a, b \in \{1,2\}$ are computed analogously.

For boundary cases where $m \in \{1,d\}$ and/or $n \in \{1,d\}$, the terms involving indices outside the range $[1,d]$ (e.g. $m - 1 = 0$) are set to zero. This approach is justified by the guard intervals (pauses) between time-frames together with the assumption that it is highly improbable for noise photons from outside the frame to jitter into the frame and cause a valid coincidence.
\subsection{Normalizing the jittered click matrices}
These resulting `jittered click matrices' represent the expected experimental data under the influence of noise, loss, and jitter. Therefore, we use the basis-normalized variants of these matrices to reconstruct a compatible density matrix $\rho_2$. For this purpose, we first normalize the jittered click matrices as stipulated by their affiliation with the theoretical bases $\mathcal{B}_0, \mathcal{B}_1$, or $\mathcal{B}_2$ by dividing them by the factors $\mathrm{CM0}, \mathrm{CM1}$ or $\mathrm{CM2}$, respectively (see Appendix \ref{app:normalization} and also \cite{BergKan}). Elements not belonging to these bases are not needed and discarded. This procedure yields the basis-normalized click matrices $\widetilde{cc}, \widetilde{cd}_{1}, \widetilde{cd}_2, \widetilde{dc}_1, \widetilde{dc}_2$ and $\widetilde{dd}_{11}, \widetilde{dd}_{12}, \widetilde{dd}_{21}, \widetilde{dd}_{22}$ used for reconstruction of $\rho_2$ as explained in Section \ref{sec:reconstruct_density}.

\section{Reconstruction of the density matrix}\label{sec:reconstruct_density}
In this section we present how to simulate and reconstruct the density matrix $\rho_2$ compatible with the jittered basis-click matrices (see Figure \ref{fig:modeling_outline}) that yields a lower bound on the key rate. Since the jittered basis-click matrices represent data that would be accessible from an experiment, the reconstruction process described here is directly applicable to empirical measurements. Importantly, apart from the diagonal, we reconstruct only lower bounds on (the real parts of) selected off-diagonal entries rather than exact values. Despite relying on different methods to evaluate the key rate, EoF and fidelity, lower bounds suffice to subsequently establish lower bounds on all three quantities.

First, the diagonal is reconstructed and (the real parts of) the first off-diagonals are lower-bounded, see Sec. \ref{subsec:first_offdiag_reconstruct}. Next, lower bounds on (the real parts of) further off-diagonal elements of interest are reconstructed using an SDP working with $3\times 3$-submatrices, see Sec. \ref{subsec:matrix_completion_offdiagonals}. We prove that this approach yields valid lower bounds which in turn can be used to attain a conservative lower bound on the key rate.\\
Before, we note that focusing on only the real parts is justified in Section \ref{subsubsec:focus_real_part} for the key rate, in Sections \ref{subsec:two_EoF_strategies} and \ref{subsubsec:focus_real_part} for the two EoF approaches and in Section \ref{subsec:valid_lower} together with Appendix \ref{app:Schmidt-bound-proof} for the fidelity.

\subsection{Direct reconstruction of diagonal and first off-diagonal}\label{subsec:first_offdiag_reconstruct}
Here, we detail how the diagonal is reconstructed and bounds on (the real parts of) certain off-diagonal elements can be found.

\subsubsection{Reconstruction of the diagonal}
As follows from the considerations in \cite[see Section IV]{BergKan}, 
the diagonal of $\rho_2$ (which of course must be real) corresponds directly to all the elements of the basis-normalized jittered TOA-TOA click matrix $\widetilde{cc}$. Thus, the diagonal of the density matrix is given by\footnote{Note that $\rho_2$ is a $d^2 \times d^2 -$ matrix, while $\widetilde{cc}$ is a $d \times d -$ matrix. Therefore one has to be careful with the indices in the correspondence between elements of $\rho_2$ and $\widetilde{cc}$.}
\begin{equation*}
        \rho_2((i-1)d + j, (i-1)d + j) := \widetilde{cc}(i,j)\text{ for } i,j = 1, \dotsc, d.
\end{equation*} 

\subsubsection{Lower-bounding of the first off-diagonals}\label{subsubsec:lower_bounding_firsts}
Next, we aim to reconstruct (the real parts of) specific off-diagonal-elements, namely those of the form $\bra{jj|\rho_2}\ket{j+1,j+1}$ and $\bra{j+1,j+1|\rho_2}\ket{jj}$ for $j=0, \dotsc, d-2$, which we name `first off-diagonal elements'.

To this end, we observe that for any density matrix $\rho$ it holds
\begin{widetext}
\begin{equation}
\begin{aligned}
    &\bra{i+,j+|\rho}\ket{i+,j+} - \ \bra{i+,j-|\rho}\ket{i+,j-} - \bra{i-,j+|\rho}\ket{i-,j+} + \bra{i-,j-|\rho}\ket{i-,j-}\\ 
    = &(\bra{i,j|\rho}\ket{i-1,j-1} + \bra{i-1,j|\rho}\ket{i,j-1} + \bra{i,j-1|\rho}\ket{i-1,j} + \bra{i-1,j-1|\rho}\ket{i,j})\\
    = &2 \text{Re}\bra{i-1,j-1|\rho}\ket{i,j} + 2\text{Re}\bra{i-1,j|\rho}\ket{i,j-1},
\end{aligned}
\end{equation}
where $\ket{l\pm}:=\frac{\ket{l}\pm \ket{l-1}}{\sqrt{2}}, \; l\in \{1, \dotsc, d-1\}$.

Noting that 
\begin{equation*}
    \begin{aligned}
        \bra{i-1,j|\rho}\ket{i,j-1} \leq |\bra{i-1,j|\rho}\ket{i,j-1}| 
    \end{aligned}
\end{equation*} 
and applying Cauchy-Schwarz to the right-hand side, 
\begin{equation*}
\begin{aligned}
    |\bra{i-1,j|\rho}\ket{i,j-1}| \leq \sqrt{\bra{i-1,j|\rho}\ket{i-1,j}\bra{i,j-1|\rho}\ket{i,j-1}} 
\end{aligned}
\end{equation*}
we arrive at 
\begin{equation*}
\begin{aligned}
    &\widetilde{dd}_{11}(i+1,j+1) - \widetilde{dd}_{12}(i+1,j+1) - \widetilde{dd}_{21}(i+1,j+1) + \widetilde{dd}_{22}(i+1,j+1)\\
    = &\bra{i+,j+|\rho}\ket{i+,j+} - \ \bra{i+,j-|\rho}\ket{i+,j-} - \bra{i-,j+|\rho}\ket{i-,j+} + \bra{i-,j-|\rho}\ket{i-,j-}\\ 
    \leq &2 \text{Re}\bra{i-1,j-1|\rho}\ket{i,j} + 2\sqrt{\bra{i-1,j|\rho}\ket{i-1,j}\bra{i,j-1|\rho}\ket{i,j-1}} \text{ for } i,j = 1, \dotsc, d-1
\end{aligned}
\end{equation*}
\end{widetext}
so that we have the bound
\begin{widetext}
\begin{equation*}
\begin{aligned}
    \text{Re}\bra{i-1,j-1|\rho}\ket{i,j} \geq \frac{1}{2}\left(\widetilde{dd}_{11}(i+1,j+1) - \widetilde{dd}_{12}(i+1,j+1) - \widetilde{dd}_{21}(i+1,j+1)\right.\\
    \left. + \widetilde{dd}_{22}(i+1,j+1)\right) - \sqrt{\widetilde{cc}(i,j+1)\widetilde{cc}(i+1,j)}\text{ for } i,j = 1, \dotsc, d-1.
\end{aligned}
\end{equation*}
\end{widetext}
By setting $i=j$, we utilize only well-defined elements of $\widetilde{dd}$, allowing us to bound the corresponding real parts of the density matrix $\rho_2$ by
\begin{widetext}
\begin{equation}\label{equ:lower_bound_first}
\begin{aligned}
    &\text{Re}\bra{k-1,k-1|\rho_2}\ket{k,k} = \text{Re}(\rho_2((k-1)d+k,kd+k+1)) \geq \frac{1}{2}\left(\widetilde{dd}_{11}(k+1,k+1) - \widetilde{dd}_{12}(k+1,k+1)\right.\\
    &\left. - \widetilde{dd}_{21}(k+1,k+1)+ \widetilde{dd}_{22}(k+1,k+1)\right) - \sqrt{\widetilde{cc}(k,k+1)\widetilde{cc}(k+1,k)}
\end{aligned}
\end{equation}
\end{widetext}
for $k = 1, \dotsc, d-1$.\\

It should be noted that the sequential matrix completion SDP relies on non-negative elements - consequently, negative first off-diagonals might yield invalid results. Therefore, we restrict our analysis to bin sizes that ensure non-negative off-diagonals. Because our simulations indicate that negative first off-diagonal values occur only for severely suboptimal bin sizes or in unrealistic regimes with extreme noise and loss, this constraint is largely theoretical and does not compromise the search for the experimental optimum.

\subsection{Matrix completion SDP for lower-bounding further off-diagonal elements}\label{subsec:matrix_completion_offdiagonals}
In this section, we present the bounding of real parts of further off-diagonal elements. We break this down into the following steps:
\begin{enumerate}
\item First, we present the SDP used for lower-bounding the non-negative real parts of further off-diagonal elements.
\item Next, we argue that sequentially applying this SDP indeed yields valid lower bounds.
\item Then, we justify the abort condition and setting concerned elements to zero.
\item Finally, we lay out why it suffices to focus only on the real parts of the matrix elements. 
\end{enumerate}

\subsubsection{SDP for sequential lower-bounding of the non-negative real parts of further elements}\label{subsubsec:further_elements}
Following the approach suggested in \cite{Tiranov_2017}, we reconstruct (the real parts of) density matrix elements farther from the diagonal by exploiting the fact that (the real part of) every $3 \times 3$-submatrix of (the real part of) any density matrix $\rho$ must be positive semi-definite.

Specifically, we aim to derive lower bounds for the real parts of elements of the form $\braket{jj|\rho}{kk},\, |j-k|\geq 2$. The reason that focusing on the real parts is sufficient lies within the way the estimated density matrix is utilized in the key rate calculation, as explained in Section \ref{subsubsec:focus_real_part}.

We initialize the procedure at the top-left of the density matrix. To bound $\text{Re}(x):=\text{Re}(\braket{00|\rho}{22}) = \text{Re}(\braket{22|\rho}{00})$, we consider the $3 \times 3$-submatrix 
\begin{equation*}
\begin{psmallmatrix}
    \braket{00|\rho}{00} & \braket{00|\rho}{11} & x\\
    \braket{11|\rho}{00} & \braket{11|\rho}{11} & \braket{11|\rho}{22} \\
    x^{\ast} & \braket{22|\rho}{11} & \braket{22|\rho}{22}
\end{psmallmatrix} =: M,
\end{equation*}
since except for $x$ and its conjugate $x^{\ast}$, all (the real parts of) its elements (or lower bounds thereof) are known. We employ an SDP that defines this $3 \times 3$-matrix as a positive semi-definite Hermitian matrix variable and has the minimization of $x + {x}^{\ast}$ as the objective.\footnote{Equivalently, one may choose $\text{Re}(x)$ or $\frac{x + {x}^{\ast}}{2}$ as objective function. In contrast, defining the matrix as real instead of complex is not equivalent as it eliminates the imaginary parts as degrees of freedom, shrinking the feasible set and potentially yielding a worse optimum and consequently non-reliable lower bounds on our figures of merit.} We demand that it has the exact same values on the diagonal as the $3\times 3$-submatrix we are using. However, the values on the first off-diagonal are not fixed, but the ones from the $3\times 3$-submatrix serve only as lower bounds, i.e. they are implemented as inequality constraints, as they were themselves lower-bounded before. Once $x$ is determined, it is saved as the corresponding entry in our estimated density matrix and may serve as a constraint for subsequent $3 \times 3$-submatrices in an SDP in the same way. By continuing this procedure, we successively determine lower-bounds on (the real parts of) off-diagonals of the density matrix using a sequence of SDPs.

Notably, we are interested only in solutions where $x$ is non-negative, which means that we impose the abort condition to stop the sequential SDPs when one estimated value is negative and set this value, the off-diagonal it belongs to, and all other remaining values that we have not lower-bounded yet to zero. The reason is that our matrix estimation technique is only valid as long as the matrix entries are non-negative and that we can justify the setting to zero of negative elements because of the way the key rate estimation works, as discussed in Section \ref{subsubsec:further_elements}.

Summing up, we use the following SDP: 
\begin{align}
\underset{M \in \mathbb{C}^{3 \times 3}}{\text{minimize}} \quad & M_{13} + M_{31}\\
\text{subject to} \quad & M \succeq 0, \label{eq:sdp_psd}\\
& M_{11} = \braket{jj|\rho}{jj}, \label{eq:sdp_diag1}\\
& M_{22} = \braket{mm|\rho}{mm}, \label{eq:sdp_diag2}\\
& M_{33} = \braket{kk|\rho}{kk}, \label{eq:sdp_diag3}\\
& \text{Re}(M_{12}) \ge L_{jj, mm}, \label{eq:sdp_ineq1}\\
& \text{Re}(M_{23}) \ge L_{mm, kk},\\
\text{return } &c^{(m)}:= \text{Re}(M_{13}),
\end{align}
whereby $M \in \mathbb{C}^{3 \times 3}$ is a complex Hermitian matrix variable representing the $3 \times 3$-submatrix formed by the states $\{\ket{jj}, \ket{mm}, \ket{kk}\}$ and $L_{jj,mm}, L_{mm,kk}$ are the best available bounds on $\text{Re}(M_{12}), \text{Re}(M_{23})$ obtained as detailed in Section \ref{subsubsec:lower_bounding_firsts} or calculated in previous iteration steps.

For a fixed pair of indices $(j, k)$, there may exist multiple intermediate indices $m$ such that $j < m < k$. To obtain the tightest possible lower bound $L_{jj,kk}$, we evaluate an SDP for each allowed intermediate index $m$ and select the maximum estimated value:
\begin{equation}
L_{jj,kk} := \max_{ j < m < k} \; c^{(m)}.
\end{equation}

\subsubsection{Validation of the sequential lower-bounding technique}
To ensure the validity of our sequential technique, we must verify that using a (non-negative) lower bound on real parts of the density matrix nearer to the diagonal we indeed also obtain valid (non-negative) lower bounds on successive real parts farther away from the diagonal and that it cannot happen that if we used a larger value for the lower bound on real parts nearer to the diagonal we would get smaller lower bounds on the elements farther away, because in this case, our sequential-SDP-technique would not yield a lower-bounded density matrix as intended.

To this end, we show that the lower the non-negtaive bound on $\text{Re}(\braket{00|\rho}{11}) = \text{Re}(\braket{11|\rho}{00}) = :y$ (and, following the same logic, $\tilde{y}:=\text{Re}(\braket{11|\rho}{22}) = \text{Re}(\braket{22|\rho}{11})$ instead of $y$) is, the lower also the non-negative bound on $x$ does get. The same then holds for any $3 \times 3$-submatrix and its off-diagonals.

The objective function is $\underset{\rho}{\text{min}}\,x + x^{\ast}$. Our conditions are that $\rho$ must be positive semi-definite (i.e. $\rho \geq 0$) and that its diagonal is equal to the one of $M$.

Naturally, without specifying any constraints, due to $\text{Re}(y) \leq |y|$ and $|\bra{i|\rho}\ket{j}|\leq \sqrt{\bra{i|\rho}\ket{i} \bra{j|\rho}\ket{j}}$ together with the fact that the diagonal of a density matrix sums to one and has only positive entries, we see that the radicand can be at most $\frac{1}{2}$, so we obtain $-\frac{1}{2} \leq \text{Re}(y) \leq \frac{1}{2}$. This means that this constraint is already implicitly contained in the SDP without us specifically stating it. Now if we demand $y$ to be lower-bounded by an element greater than $-\frac{1}{2}$, this further restricts our solution space, so that the minimum we find for the objective function with this further restriction cannot be lower than the minimum we find for the objective function without it, since we have fewer possible choices for $\rho$. But as we are using a valid lower bound to constrain $y$, we don't restrict the solution space for $x$ too much, such that what we get by minimizing $x + x^{\ast}$ yields a lower bound on $\text{Re}(x)$ and cannot overestimate it.

Therefore, by using a lower bound on the element $\text{Re}(y)$, the solution we find for the minimization of $\text{Re}(x)$ can also only be a lower bound on $\text{Re}(x)$ (and can never overestimate it as the bound we would get for $\text{Re}(x)$ when using a constraint with a larger lower bound on $\text{Re}(y)$ could only yield a larger (or equal) solution to the minimization problem of $\text{Re}(x)$). If we then proceed our consecutive SDP with the lower bound on $\text{Re}(x)$, we can only get a lower bound on the missing offidagonal-element in the next $3\times3$-submatrix and so on, so that we end up with a density matrix the estimated off-diagonal-elements of which indeed contain lower-bounds on the real parts of the elements of any density matrix compatible with the experimental data.

Thus, by using the known and the estimated elements of $\rho_2$ for the key rate calculation, we obtain a lower bound on the key rate, because the calculation is implemented in a way such that the lower those non-negative real parts are, the lower the estimated key rate is (see \cite{KanDual} and its supplemental material).

\subsubsection{Justification of the abort condition}
As noted in Section \ref{subsubsec:further_elements}, we terminate the matrix completion as soon as a lower bound is estimated to be negative, set the corresponding off-diagonal element (and the ones in the same diagonal) to zero, and retain only the positive bounds calculated up to that point for the key rate and in the second approach for the EoF.\footnote{Note that the completed matrix is neither used in the first approach for the EoF nor in the bounding of the fidelity, since both methods use their own SDP which depends only on the diagonal and first off-diagonal elements.} We now justify the validity of this procedure for both applications.
\begin{itemize}
\item Key rate (see Section \ref{sec:lower_bound_key}): The reason for the validity lies within the method we utilize for the key rate calculation: Following \cite{KanDual}, we use the partly completed matrix $\rho_2$ only in the form $\text{Tr}(\rho_2 \hat{W})$, where $\hat{W}$ is some observable we can choose and $\text{Tr}(.)$ denotes the trace operator. Therefore, setting elements in $\rho$ to zero can be understood as setting certain entries of $\hat{W}$ to zero (namely the element $\hat{W}_{ij}$ if $\rho_{ji}$ was set to zero).
\item Second EoF approach (see Section \ref{subsec:two_EoF_strategies}): The sequential SDP only determines off-diagonal values, which appear in the formula for the bound on the EoF, Equ. \eqref{equ:C_in_EoF_bound}, only in the form $|\bra{jj|\rho}\ket{kk}|$, $j>k$. The larger those values are, the larger also the EoF will be. Since it trivially holds that $0 \leq |\bra{jj|\rho}\ket{kk}|$, we simply can bound these values by zero.
\end{itemize}

\subsubsection{Validation of focusing on the real part}\label{subsubsec:focus_real_part}
Lastly, we want to justify that we can work with $\text{Re}(\rho)$ instead of $\rho$ in the key rate calculation. Here, the crucial aspect also lies within the choice of the observable $\hat{W}$, which is described in the supplemental material of \cite{KanDual}\footnote{\url{http://link.aps.org/supplemental/10.1103/PhysRevLett.135.010802}}: We only use observables which are symmetric, which means that for these we have $\text{Tr}(\rho \hat{W}) = \text{Tr}(\text{Re}(\rho)\hat{W})$. 
Therefore, the estimates on the real part of the matrix computed with our matrix completion technique are legitimate to use for the subsequent key rate calculation.

Though we make use of a different method for the bounding of the density matrix we use for the EoF calculation as it yields superior lower bounds on the EoF, for dimensions higher than about $64$ that technique is not feasible due to limited computational power. Therefore, we briefly explain why it is justified to use the matrix calculated with the sequential technique as input to the EoF calculation:

The diagonal of the density matrix is fixed and our SDP only determines off-diagonal values, which appear in the formula for the bound on the EoF, Equ. \eqref{equ:C_in_EoF_bound}, only in the form $|\bra{jj|\rho}\ket{kk}|$, $j>k$. The larger this value is, the larger also the EoF will be. Since we are aiming to lower-bound the EoF and it trivially holds that $|\bra{jj|\text{Re}(\rho)}\ket{kk}| \leq |\bra{jj|\rho}\ket{kk}|$, from which follows
\begin{equation}
    \text{EoF}(\rho) \geq -\text{log}_2(1-\frac{(B(\rho))^2}{2}) \geq -\text{log}_2(1-\frac{(B(\text{Re}(\rho)))^2}{2}),
\end{equation}
 we can use lower bounds on the absolute values of the elements of $\text{Re}(\rho)$ and set the imaginary part to zero. Such lower bounds are exactly what we get from our sequential matrix completion, since there we lower-bound the real part of $\rho$ as long as the bound is positive and set it to zero if it gets negative, which is trivially justified due to $0 \leq |\bra{jj|\text{Re}(\rho)}\ket{kk}| \leq |\bra{jj|\rho}\ket{kk}|$.

\section{Lower-bounding the key rate}\label{sec:lower_bound_key}
Here, we explain how to use the simulated quantities in order to obtain a valid lower bound on the key rate.

We determine the lower bound on the secure key rate following the framework for high-dimensional entanglement-based protocols detailed in \cite{BergKan}.
As discussed in Section \ref{subsec:key_theory}, we employ the Devetak-Winter bound to lower-bound the key rate per coincidence click, utilizing the standard formulation, which can be obtained from 
\begin{equation*}
    r_{\text{cc}} \geq I(A:B)_{\text{cc}} - I(A:E)_{\text{cc}}
\end{equation*}
by using the definition of $I(X:Y)$ being the difference between the entropy of $X$ and the conditional entropy of $X$ given $Y$, $I(X:Y):=H(X)-H(X|Y)$, yielding
\begin{equation}
 r_{\text{cc}} \geq S(A|E)_{\text{cc}}-H(A|B)_{\text{cc}},   
\end{equation}
where $S(A|E)_{\text{cc}}$ is the conditional von Neumann entropy of Alice's key register given Eve's quantum register, quantifying Eve's lack of knowledge about Alice's bit string $A$, and $H(A|B)_{\text{cc}}$ is the Shannon entropy between Alice's and Bob's raw key, representing the amount of information that Alice needs to communicate to Bob to reconcile their keys. The latter term is classical and accessible from the observed statistics (see Appendix \ref{app:ecterm}), but to lower-bound the asymptotic secure key rate, we have to lower-bound the first term. To do that, we use the min-entropy $H_{\text{min}}(A|E)_{\text{cc}}$, which can be obtained from $p_{\text{guess}}$, the probability that Eve guesses Alice's bit string correctly, $H_{\text{min}}(A|E)_{\text{cc}} = -\text{log}_2(p_{\text{guess}})$.

To calculate the guessing probability and subsequently obtain the lower bound on the key rate, we incorporate the semi-analytic method suggested in \cite{KanDual} under usage of the already reconstructed elements of $\rho_2$ (see Section \ref{sec:reconstruct_density}) and observables as introduced in \cite{KanDual}, which were inspired by entanglement witness theory \cite{Friis_2018}).

The thereby obtained lower bound on the key rate per coincidence click, $r_{\text{cc}}$, can be converted into a lower bound on the key rate per second via the formula 
\begin{equation}
    r_{\text{sec}} = r_{\text{cc}}P_{\text{click}} p_{\text{CC}} n_{\text{frames}},
\end{equation}
where $r_{\text{sec}}$ denotes the key rate per second, $n_{\text{frames}}$ the number of frames per second (with an inter-frame pause in the length of one time-bin), $P_{\text{click}}$ the probability that exactly one coincidence click occurs within a frame and $p_{\text{CC}}$ the probability of a TOA-TOA-measurement, as detailed in appendix \ref{app:keyrate_per_second}.

\section{Lower-bounding the EoF}\label{sec:lower-bounding_EoF}
Next, we turn to the task of lower-bounding the Entanglement of Formation. As suggested in \cite{Martin_2017}, we use an SDP for handling the unknown off-diagonal elements. However, one has to be careful here in order to obtain a genuine lower bound, as detailed below.

\subsection{Strategies for lower-bounding the EoF}
The first strategy involves selecting a set of index pairs C a priori to define the objective function of the SDP. This SDP identifies the `worst-case' density matrix—the one that minimizes $B \geq 0$ (introduced in Equ. \eqref{equ:EoF_bound} and defined in Equ. \eqref{equ:C_in_EoF_bound}) among all states compatible with the experimental data—thereby providing a safe lower bound on the EoF. Since any such set C yields a mathematically valid lower bound, the ultimate goal is to identify the specific set C that results in the maximal (tightest) lower bound.

The second approach is to first reconstruct the density matrix $\rho$ by lower-bounding the modulus of each off-diagonal element of the form $|\bra{jj|\rho}\ket{kk}|$ in advance (e.g. under usage of the sequential matrix completion technique described in Section \ref{sec:reconstruct_density}) and then determine the set $C$ which yields the best EoF-bound for this resulting density matrix.

As intuitively expected, when one applies the first method several times (namely for various sets $C$) and chooses the best of it, this generally yields a tighter bound than the second approach. However, because the first approach processes the entire density matrix (or at least its reduced submatrix as explained below) simultaneously, its computational complexity grows rapidly. In contrast, the second approach - utilizing sequential $3 \times 3$-submatrix SDPs - is significantly more scalable and can still be used for high-dimensional systems for which the first approach becomes computationally intractable. 

\subsection{Details on the two strategies for lower-bounding the EoF}\label{subsec:two_EoF_strategies}
More precisely, the two approaches look the following: 
\begin{enumerate}
\item The first approach is to calculate lower bounds on the EoF for various sets $C$ and their respective worst-case density matrix compatible with the constraints and to then take the tightest bound among these. To this end, for every set $C$ (defined below Equ. \eqref{equ:C_in_EoF_bound}), $B(\rho)$ (defined in Equ. \eqref{equ:C_in_EoF_bound}) has to be minimized, since the smaller $B(\rho) \in [0, 1)$, the smaller the bound on the EoF (in Equ. \eqref{equ:EoF_bound}) is.
\begin{enumerate}
        \item First approach, Full-state-version: Noting that Equ. \eqref{equ:C_in_EoF_bound} can be rewritten as 
        \begin{widetext}
\begin{equation}\label{equ:C_alternatively}
        B(\rho) = \frac{2}{\sqrt{|C|}}\sum\limits_{\substack{(j,k)\in C \\ j < k}} |\bra{jj|\rho}\ket{kk}| - \frac{2}{\sqrt{|C|}}\sum\limits_{\substack{(j,k)\in C \\ j < k}}\sqrt{\bra{jk|\rho}\ket{jk}\bra{kj|\rho}\ket{kj}},
\end{equation} \end{widetext} and that only the first summand on the right contains variable elements, while the elements in the second sum are already fixed as they are diagonal elements, it becomes clear that the worst-case density matrix for $B(\rho)$ is the one which minimizes the first sum and is still compatible with our constraints. This yields the following procedure: First, the set $C$ from Equ. \eqref{equ:C_alternatively} is fixed. Then one defines the SDP variable $\rho_{\text{var}}$ as a Hermitian positive semi-definite $d^2 \times d^2$-matrix and incorporates the constraints as follows:
 \begin{widetext}   \begin{align}
\underset{\rho_{\text{var}} \in \mathbb{C}^{d^2 \times d^2}}{\text{minimize}} \quad & \sum_{\substack{(j,k)\in C \\ j < k}} \left| \text{Re}\big(\bra{jj}\rho_{\text{var}}\ket{kk}\big) \right|\\
\text{subject to} \quad & \rho_{\text{var}} \succeq 0, \label{eq:sdp_eof_psd}\\
& \bra{ij}\rho_{\text{var}}\ket{ij} = \bra{ij}\rho_2\ket{ij}, \quad \forall \, i,j \in \{0, \dotsc, d-1\}, \label{equ:equ_first}\\
&\text{Re}\big(\bra{jj}\rho_{\text{var}}\ket{j+1,j+1}\big) \ge \text{Re}\big(\bra{jj}\rho_2\ket{j+1,j+1}\big), \quad \forall \, j \in \{1, \dotsc, d-1\}. \label{equ:inequ_first_off} 
\end{align}
\end{widetext} 
Therein, the physical knowledge of the state is accounted for through the equality constraints for the diagonal (Equ. \eqref{equ:equ_first}) and the inequality constraints for the first off-diagonal (Equ. \eqref{equ:inequ_first_off}),
    using the bounds obtained as described in Section \ref{subsec:first_offdiag_reconstruct}. Since the objective function is monotonically non-decreasing with respect to these real parts, utilizing a non-negative lower bound\footnote{Remember that the technique must not be used if the bounded first off-diagonals contain negative elements.} as an inequality constraint maintains the validity of the final lower bound on the EoF. Note that a separate constraint for the conjugate elements is redundant due to the Hermitian property of $\rho_{\text{var}}$.
        Since the resulting EoF bound depends on the choice of $C$, we run the SDP for several different sets and select the one that yields the highest lower bound for its corresponding worst-case density matrix.
        In our calculations, we explore $d-1$ different sets for $C$, where we start with $C_1:=\{(1,2), (2,3), (3,4), \dotsc, (d-1,d)\}$, and progressively expand it: $C_2:=C_1 \cup \{(1,3), (2,4), (3,5), \dotsc, (d-2,d)\}$ and so on.
             
        \item First approach, Reduced-state-version: As a more efficient alternative to the full-state approach, we propose a modification that significantly reduces computational overhead by projecting the problem onto a relevant subspace. Instead of working with the full $d^2 \times d^2$-matrix, one extracts the relevant entries into a $d\times d$-matrix $\rho_{\text{red}}$, defined as 
        \begin{widetext}
        \begin{equation}
            \rho_{\text{red}}:=
            \begin{psmallmatrix}
                \bra{00|\rho_2}\ket{00} & \bra{00|\rho_2}\ket{11} & \bra{00|\rho_2}\ket{22} & \hdots & \bra{00|\rho_2}\ket{d-1,d-1}\\
                \bra{11|\rho_2}\ket{00} & \bra{11|\rho_2}\ket{11} & \bra{11|\rho_2}\ket{22} & \hdots & \bra{11|\rho_2}\ket{d-1,d-1}\\
                \vdots & \vdots & \vdots & \ddots & \vdots\\
                \bra{d-1,d-1|\rho_2}\ket{00} & \bra{d-1,d-1|\rho_2}\ket{11} & \bra{d-1,d-1|\rho_2}\ket{22} & \hdots & \bra{d-1,d-1|\rho_2}\ket{d-1,d-1}
            \end{psmallmatrix}.
        \end{equation}
        \end{widetext}
        
        One then works with an SDP in analogy to the one described for the full-state-version, where this time $\rho_{\text{var}}$ is not a $d^2 \times d^2$-, but a $d \times d$-matrix. This has the advantage that the computation is still feasible for large dimensions in which the full-state approach fails due to limited computational power.
        
        Theoretically, focusing on a $d \times d$-subspace is less restrictive than optimizing over the full space, which should nominally result in looser EoF bounds. However, our numerical tests indicate that the difference is negligible. (Interestingly, we occasionally observe slightly higher results with the reduced version, contrary to theoretical expectations, which we attribute to numerical solver tolerances within the Matlab cvx toolbox.) For the simulations presented in this work, we utilize the full-state-version.
    \end{enumerate}
    \item [] It should be emphasized that one must not attempt to reconstruct the off-diagonal-elements and optimize the set $C$ simultaneously, as this could compromise the rigor of the lower bound and lead to an overestimation of the EoF.
    \item [] We note that the legitimacy of using a lower bound on only the real parts of the first off-diagonals lies within the form of $B$ in Equation \ref{equ:C_alternatively}: Since 
    \begin{equation}\text{Re}(\bra{jj|\rho}\ket{kk})\leq |\text{Re}(\bra{jj|\rho}\ket{kk})| \leq |\bra{jj|\rho}\ket{kk}| \, \forall j,k,
    \end{equation}
    any non-negative lower bound on the first off-diagonal elements $\text{Re}(\bra{jj|\rho}\ket{j+1,j+1}), \, j=1, \dotsc, d-1$ does result in a lower bound on the first sum in Equation \ref{equ:C_alternatively}, which leads to a lower bound on $B$ and in turn on the EoF, since it is monotonically increasing in $B$. Furthermore, in the SDP (see Equ. \ref{eq:sdp_eof_psd} to \ref{equ:inequ_first_off}) only the real parts are used.
    
    \item Second approach: An alternative approach is to make use of the estimation of $\rho_2$ obtained through the sequential matrix completion described in Section \ref{sec:reconstruct_density}. As noted there previously, and as suggested already in \cite{Tiranov_2017}, using that estimation is a valid method for determining the EoF. Since in this approach the density matrix is already fixed, the objective shifts to optimizing the set $C$ from Equ. \eqref{equ:C_in_EoF_bound}) to maximize $B$.
    
    We do this by starting with $C=\{\}$ and letting $(k,l)$ run through the pairs of indices from $(1,1), (1,2), \dotsc, (1,d)$ over $(2,3), (2,4), \dotsc, (2,d) \dotsc$, up to $(d-1,d)$, where each pair $(k,l)$ of these pairs of indices is added to $C$ whenever  \begin{widetext}{\begin{equation*}\resizebox{0.95\textwidth}{!}{$\left(\frac{2}{\sqrt{|C\cup\{(k,l)\}|}}\sum\limits_{\substack{(m,n)\in C\cup\{(k,l)\}\\m < n}}\left( |\langle mm|\rho_2|nn\rangle|-\sqrt{\langle mn|\rho_2|mn\rangle\langle nm|\rho_2|nm\rangle}\right)\right)^2 > \left(\frac{2}{\sqrt{|C|}}\sum\limits_{\substack{(m,n)\in C\\m < n}}\left(|\langle mm|\rho_2|nn\rangle|-\sqrt{\langle mn|\rho_2|mn\rangle\langle nm|\rho_2|nm\rangle}\right)\right)^2,$}\end{equation*}}\end{widetext} in which while $C = \{\}$, the right-hand side is assumed to be zero.
    
    Finally, under usage of the resulting set $C$ the lower bound on the EoF is calculated according to Equ. \eqref{equ:EoF_bound}.
\end{enumerate}
It is important to point out that within both approaches, whenever the resulting value for $B$ is negative, it is set to zero, since it would yield a false positive lower bound on the EoF due to the squaring in Equ. \eqref{equ:EoF_bound}, as mentioned in Section \ref{subsec:EoF_theory}.

\section{Lower-bounding the Fidelity}\label{sec:lower-bounding_Schmidt}
In this section, we turn to the task of lower-bounding the fidelity. First, we present the SDP used for it, then we prove the validity of the bound. Lastly, we present an alternative approach for handling higher dimensions efficiently.

\subsection{Calculation of the lower bound}
To determine the lower bound on the fidelity, we construct an SDP based on the inequality presented in Section \ref{subsec:Schmidt_theory}, following a procedure similar to the first approach for lower-bounding the EoF.

More precisely, the objective function to be minimized is the fidelity $\mathcal{F}(\rho_{\text{var}}, \ket{\Psi_T})$ (defined in Equ. \eqref{equ:fidelity_definition}) of the Hermitian semi-definite $d^2 \times d^2$-variable $\rho_{\text{var}}$ to the maximally entangled target state (see Section \ref{subsec:Schmidt_theory}). The constraints are identical to those employed for the EoF lower bound, utilizing the observed diagonal elements and the lower-bounded real parts of the first off-diagonals of $\rho_2$, yielding: 
\begin{widetext}   \begin{align}
\underset{\rho_{\text{var}} \in \mathbb{C}^{d^2 \times d^2}}{\text{minimize}} \quad & \mathcal{F}(\rho_{\text{var}}, \ket{\Psi_T})\\
\text{subject to} \quad & \rho_{\text{var}} \succeq 0, \\
& \bra{ij}\rho_{\text{var}}\ket{ij} = \bra{ij}\rho_2\ket{ij}, \quad \forall \, i,j \in \{0, \dotsc, d-1\},\\
& \text{Re}\big(\bra{jj}\rho_{\text{var}}\ket{j+1,j+1}\big) \ge \text{Re}\big(\bra{jj}\rho_2\ket{j+1,j+1}\big), \quad \forall \, j \in \{1, \dotsc, d-1\}.
\end{align}
\end{widetext}

Upon completion of the SDP, the minimized value of the objective function represents a lower bound for the fidelity $\mathcal{F}(\rho_2, \ket{\Psi_T})$ which in turn is a lower bound on $\mathcal{F}_{\text{max}}(\rho_2)$ (introduced in Equ. \eqref{equ:max_fidelity_definition}). Therefore, the lower bound on the Schmidt number is given by the smallest integer value $1 \leq j \leq d$ which satisfies $\frac{j-1}{d} < \mathcal{F}(\rho) \leq \frac{j}{d}$.

\subsection{Validity of the lower bound on the fidelity}\label{subsec:valid_lower}
To ensure the validity of our results, we must justify why employing a lower bound on (the real parts of) the first off-diagonals cannot cause an overestimation of the fidelity and Schmidt number. To this end, in Appendix \ref{app:Schmidt-bound-proof} we show that for any density matrix $\rho$ the relationship $\braket{\Psi_{T}|\tilde{\rho}}{\Psi_{T}} \leq \braket{\Psi_{T}|\rho}{\Psi_{T}}$ holds, where $\tilde{\rho}$ is the matrix obtained from $\rho$ by replacing the real parts of the first off-diagonal elements with their respective lower bounds. This confirms that the substitution preserves the lower-bound property of the fidelity.

Furthermore, replacing equality constraints with inequality constraints within the SDP expands the feasible region of the SDP. In any minimization problem, an expanded search space can only result in a minimum that is less than or equal to the minimum found in a more restricted space. Consequently, our approach is guaranteed to yield a conservative lower bound on the fidelity, which in turn ensures a valid lower bound on the fidelity and Schmidt number. 

\subsection{Efficiently handling very high dimensions}
It should be noted that the efficacy of the aforementioned method decreases in regimes characterized by higher dimensions or increased noise, loss, and jitter. Under these conditions, the SDP may yield reconstructed matrices with negative off-diagonal elements, which can compromise the tightness of the bound. Therefore, for dimensions starting from around $16$, we recommend utilizing the inequality 
\begin{equation} 
    \underset{\rho \in S_k}{\text{max}} \sum\limits_{i,j = 0}^{d-1}|\bra{ii|\rho}\ket{jj}| \leq k,
\end{equation}
where $S_k$ denotes the set of all density matrices with Schmidt number at most $k$, following the approach presented in \cite{Schiffer_2026}. This criterion leads to an SDP formulation similar to the one described above, albeit with a different objective function.

However, since our simulations focus on the range between $d=2$ and $d=8$, and as the fidelity and Schmidt number are not the quantities we prioritize in our calculations, we have utilized the former computational method for all results regarding these two quantities.

\section{Discussion}\label{sec:discussion}
In this section, we discuss the simulation results obtained by application of our proposed noise and jitter model. (We refer to Appendix \ref{app:realistic_simulations} for simulation details.)

To address the fundamental question of whether high-dimensional entanglement outperforms qubit-based systems in QKD, we apply the model to a variety of realistic parameter regimes. While previous experiments have indicated an advantage for higher dimensions, these setups typically involve parameters fixed \textit{a priori}. Consequently, it remained unclear whether a different configuration - such as an alternative bin size - might have shifted the maximum key rate back to the two-dimensional counterpart. Our simulations demonstrate that this is not the case: There are indeed practically relevant regimes in which dimensions higher than two yield the optimal key rate for QKD.

\subsection{Observations regarding the key rate per second}
\begin{figure*}[!htbp]
    \centering
    \includegraphics[width=1.\textwidth]{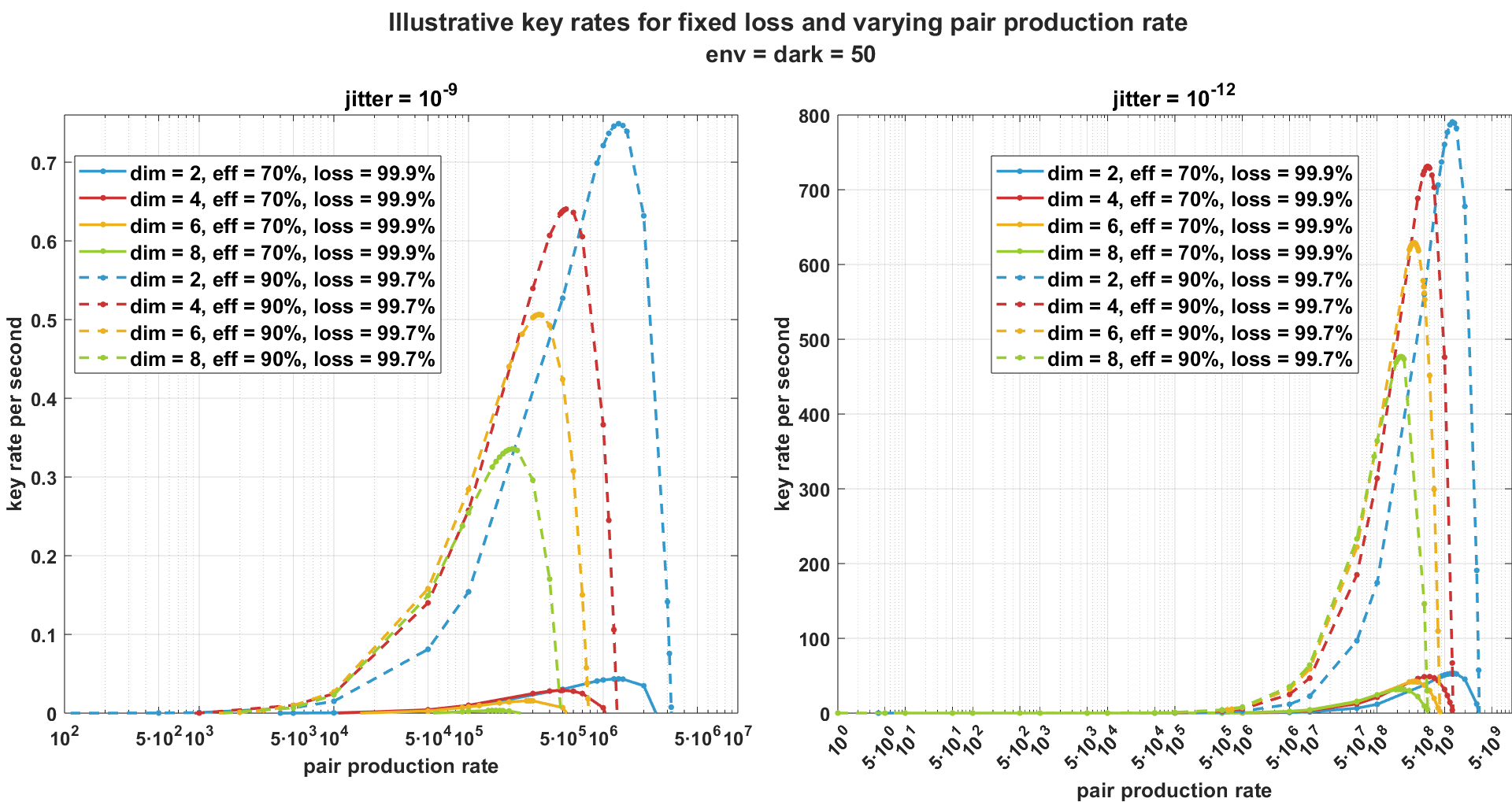}

\caption{Visualization of the key rate per second as a function of the pair production rate for dimensions $d \in \{2,4,6,8\}$. All curves are obtained using optimized bin sizes. The subplots compare different timing jitter levels: jitter-$\sigma = 10^{-9}$sec on the left and jitter-$\sigma = 10^{-12}$sec on the right. Both subplots correspond to an environmental photon rate and a dark count rate of $50$ photons per second respectively (on both Alice's and Bob's side). Each of the two shows the curves for a detector efficiency of 70\% and loss probability of 99.9\% as well as a detector efficiency of 90\% and a loss probability of 99.7\%.}
\label{fig:plot_keyrate_production_rate}
\end{figure*}

In Figure \ref{fig:plot_keyrate_production_rate} we present the key rate per second as a function of the pair production rate for various parameter regimes. A primary observation is the existence of crossover points between the curves of different dimensions. While there is always a pair production rate for which dimension two performs better than higher dimensions, the crossing illustrates that there are regimes of the pair production rate for which higher dimensions are better than lower ones. This demonstrates that the optimal choice of dimension is strictly dependent on the pair production rate achievable within a specific experimental setup.\footnote{Note that the pair production rate cannot be increased arbitrarily, but is fundamentally constrained by practical limits.} In our data analysis, we observe a nested structure in the pair production intervals that yield a positive key rate: The ones for higher dimensions are typically subsets of the ranges for lower dimensions. In other words, lower dimensions exhibit greater robustness at both very low and very high pair production rates.

Similarly, for all our simulation results it holds that for each dimension the pair production interval yielding positive key rates under lower detection efficiency and higher loss is contained within the corresponding range for higher detection efficiency and lower loss.

Comparing the left and right panels of Figure \ref{fig:plot_keyrate_production_rate} reveals the profound impact of temporal jitter. Lower jitter does not merely lead to substantially higher peak key rates, but also shifts these maxima toward considerably higher pair production rates and drastically expands the interval over which the key rate remains positive. The corresponding optimal bin sizes are significantly smaller for smaller jitter compared to higher jitter. (For plots and discussion of the optimal bin size see Section \ref{subsec:discussion_binsize}).

Additional simulations (omitted from the figure for clarity) verify the natural supposition that reducing environmental photons and dark counts further enhances performance, yielding marginally higher key rates and an expanded interval of the pair production rate for which the key rate is positive. Furthermore, the peak key rate shifts to a higher pair production rate and a smaller corresponding optimal bin size.

\begin{figure*}[!htbp]
    \centering
    \includegraphics[width=1.\textwidth]{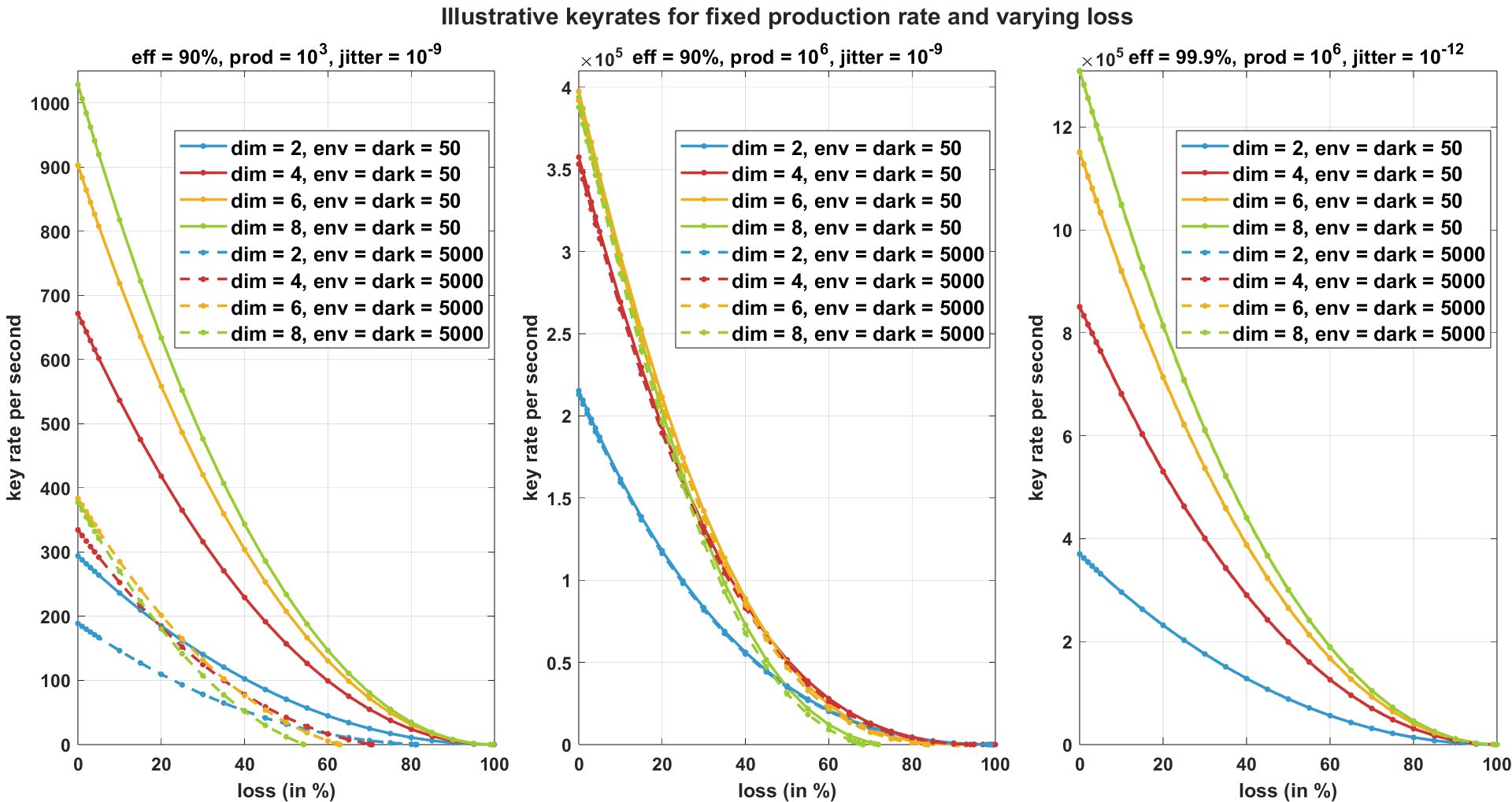}
\caption{Visualization of the key rate per second as a function of the channel loss for dimensions $d\in \{2,4,6,8\}$, achieved for optimized bin sizes. The subplots represent different parameter regimes: In the subplot on the very left, one can see the key rate for a jitter-$\sigma$ of $10^{-9}$, detector efficiency of $90\%$ and pair production rate of $10^3$ photons per second, while in the one in the middle the pair production rate is increased to $10^6$ photons per second, and in the subplot on the very right the detector efficiency is set to $99.9\%$ and the jitter-$\sigma$ to $10^{-12}$, while the pair production rate stays at $10^6$ photons per second. Each of the three subplots depicts the curves for two noise levels: an low-noise environment with an environmental photon rate and dark count rate of $50$ photons per second  and a high-noise environment with both rates increased to $5000$ photons per second. In the third subplot, both curves are so similar that they dashed one is covered by the solid one.}
\label{fig:plot_keyrate_loss}
\end{figure*}

In contrast to the previous analysis, Figure \ref{fig:plot_keyrate_loss} illustrates the key rate per second as a function of the loss probability for a fixed pair production rate across various parameter configurations. Here the maximum key rate is frequently achieved by higher-dimensional systems rather than qubits. The crossing of the dashed curves in the first and second subplot - particularly when compared to the solid lines - confirms that the optimal dimension is strictly parameter-dependent. This again validates that there are physically relevant and realistic regimes for which high dimensional entanglement provides a definitive advantage.

In our simulations with the bin size optimized in plausible regions and with fixed jitter, we observe a clear correlation: Lower jitter facilitates higher optimal dimensions, which can be explained by the fact that lower jitter allows for smaller bin sizes without introducing a strong jitter effect, whereas increased jitter favors lower-dimensional protocols. However, also in the former case, the advantage of going to higher dimensions is expected not to be infinite as even in the idealized absence of jitter, the benefit of increasing dimensions eventually diminishes as the bin size cannot be made arbitrarily small. Therefore, increasing the dimension at some point will lead to necessarily larger frames and therefore will be accompanied by a decline in frequency of valid coincidence clicks due to multi-click events where more than one detection is regarded within a single frame for at least one party (which are discarded since the protocol requires exactly one coincidence click per frame).

Although visually compressed in Figure \ref{fig:plot_keyrate_loss}, numerical determination of the intersection with the $x$-axis shows that lower-dimensional systems tolerate greater loss while still maintaining a positive key rate than their higher-dimensional counterparts. This aligns with the intuitive expectation that as loss increases, dark counts and environmental photons gain more weight, which is reflected by a decrease in the signal-to-noise ratio. Higher-dimensional systems are particularly put at disadvantage, because they are inherently associated with longer time-frames and attempting to mitigate the frame expansion by shrinking the bin size is accompanied by increased jitter effects.

Naturally, with fewer environmental photons and dark counts, the key rate stays positive at higher loss probabilitys due to an improved signal-to-noise ratio. However, the overall impact of noise on the optimal dimension is strongly governed by its magnitude relative to the pair-production rate, as evidenced by the divergence between the first two subplots in Figure \ref{fig:plot_keyrate_loss}.

\subsection{Observations regarding the optimized bin sizes}\label{subsec:discussion_binsize}
\begin{figure*}[!htbp]
    \centering
    \includegraphics[width=1.\textwidth]{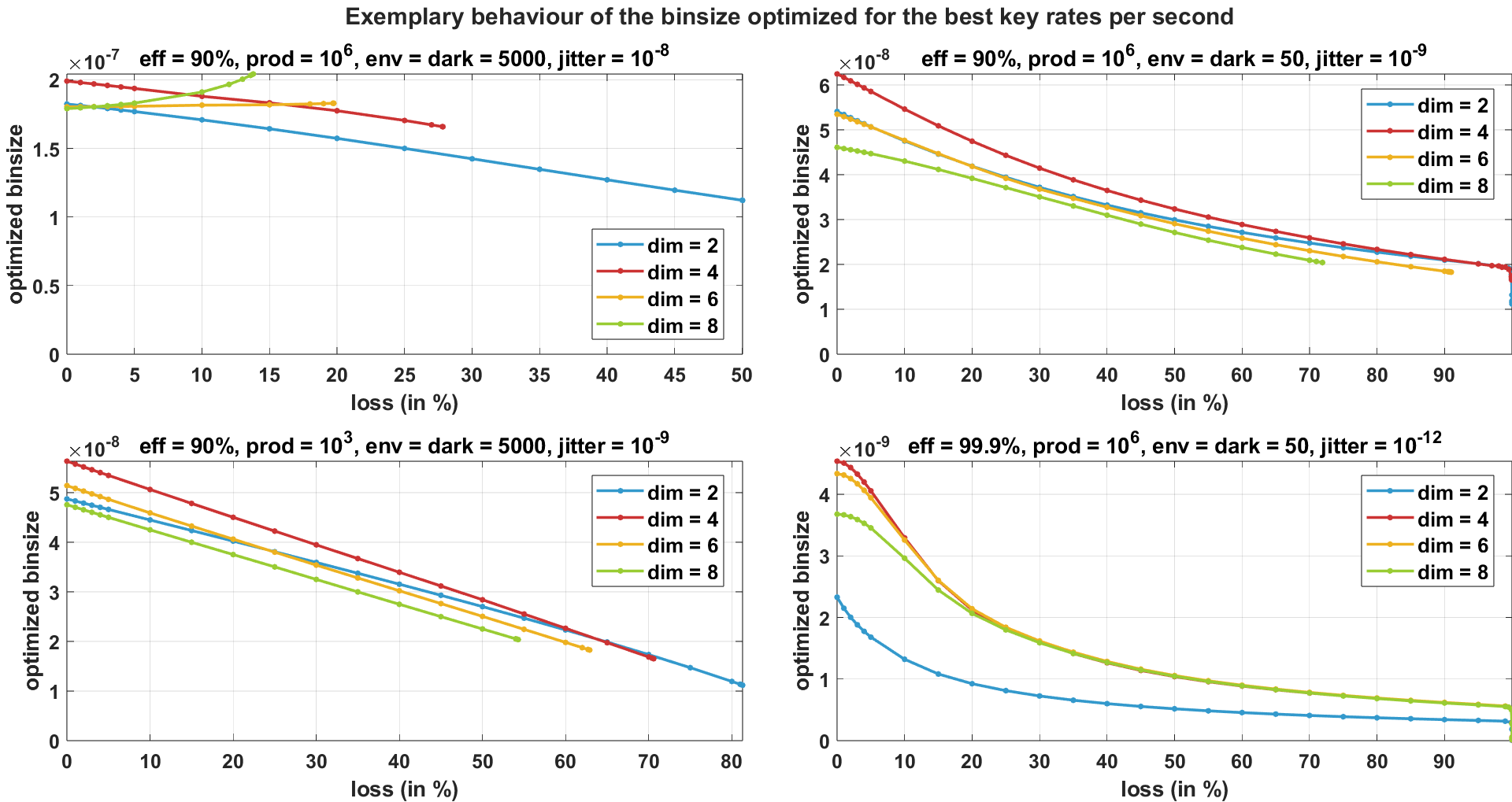}
    \caption{Visualization of the optimized bin size as a function of the channel loss: The subplots depict the bin size (in seconds) that maximizes the key rate per second for various dimensions. The results display how different it behaves and which orders of magnitude it attains for fixed pair production rate and varying loss for illustrative parameter configurations.}
    \label{fig:plot_binsizes}
\end{figure*}

As illustrated in Figure \ref{fig:plot_binsizes}, the bin size required to maximize the key rate per second exhibits a non-trivial dependency on the underlying physical parameters.

For a fixed pair production rate, the optimal bin size generally attains its maximum at zero loss and decreases as the loss increases. A notable exception occurs at a higher jitter (jitter-$\sigma=10^{-8}$, where the optimal bin size decreases for dimensions two and four, but increases with loss for dimensions six and eight.

In contrast, an examination of the simulated data reveals that when varying the pair production rate at a fixed loss, the bin size follows a non-monotonic trend: It starts off small, increases with increasing pair production rate, reaches a peak and decreases again. Crucially, for each dimension, the peak coincides with the pair production value that also maximizes the metrics per coincidence click (i.e. EoF, key rate and fidelity). While these quantities attain their maximum at the same pair production rate within a single dimension, the specific dimension that yields the global maximum differs across these various metrics (as well as across the entanglement rate and key rate per second). More details are given in Figures \ref{fig:prod_table} and \ref{fig:loss_table} in Appendix \ref{app:detailed_results}.

Our results indicate that the order of magnitude of the optimal bin size is primarily determined by the timing jitter and the pair production rate.  Furthermore, we observe that a decrease in detector efficiency is consistently accompanied by a reduction in the optimal bin size. This can be attributed to the changing weight of different noise sources. A lower detection efficiency reduces the signal (and environmental photon) rate, thus increasing the relative impact of dark counts. On the other hand, a smaller bin size mitigates this by reducing the probability of a dark count occurring within a time-frame.

\subsection{Observations regarding further quantities}
While the key rate per second is the primary figure of merit for QKD applications, other quantities - such as the entanglement rate, the EoF, the fidelity, the Schmidt number, and the key rate per coincidence click - are vital for broader quantum networking tasks and therefore are included in our analysis, too.

As expected, the key rate per second attains its maximum value at a different pair production rate than the other metrics of interest, which becomes clear by realizing that - in contrast to the EoF per coincidence click, fidelity and key rate per coincidence click - the key rate per second is intrinsically linked to the absolute frequency of coincidence detections.

Furthermore, it is important to note that as the bin size is specifically optimized to maximize the key rate per second, the values obtained for the other quantities might be sub-optimal and optimizing the bin size for other metrics might yield substantially different results. To investigate this possibility, we perform a comparative analysis for two specific regimes (one with fixed loss and variable pair production rate and one with variable pair production rate and fixed loss). In these cases, the bin size was re-optimized independently for the key rate per coincidence click, for the EoF, for the fidelity and for the entanglement rate. 

\subsubsection{Discussion of the EoF, Fidelity, Schmidt number and key rate per coincidence click for bin sizes optimized for the key rate per second}\label{subsubsec:discuss_EoF_and_others}
In this section, we discuss the behavior of the EoF, fidelity, Schmidt number, and key rate per coincidence click, specifically for bin sizes optimized to maximize the key rate per second.
In our simulations with fixed loss and varying pair production rate, we observe that for any given dimension, these four quantities attain their maximum at the same pair production rate. This synchronized behavior is likely attributed to their dependence on the ratio of valid signal events (i.e. unjittered source photon pair clicks) to all other coincidence events (caused by jitter, environmental photons or dark counts). While the peaks align within a single dimension, the global maximum across all dimensions does not necessarily occur at the same dimension, as depicted in Figure \ref{fig:prod_table_main}. For further details, we refer to Figure \ref{fig:prod_table} in Appendix \ref{app:detailed_results}.
\begin{figure*}[!htbp]
    \centering
    \includegraphics[]{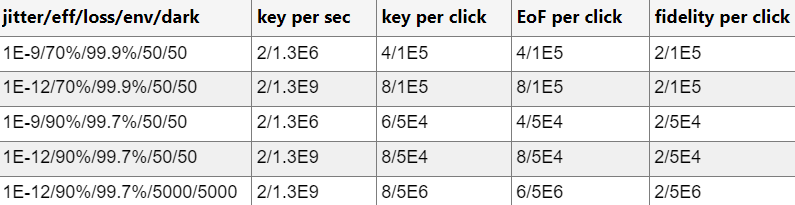}
    \caption{For five different parameter constellations (see column one), the table shows where the maximum of the quantities is achieved (see columns two to five), with the first number giving the dimension and the second one the corresponding best pair production rate (in emitted photon pairs per second) within this dimension. (e.g. For a jitter-sigma of $10^{-9}$ seconds, detection efficiency of 70\%, photon loss probability of 99.9\% and environmental photon count as well as dark count rate of 50 photons per second, the highest key rate per second - which is the quantity we optimized the bin size for - is obtained for dimension two, at a pair production rate of $1.3*10^6$ photon pairs per second, the highest key rate per coincidence click and EoF per coincidence click both occur at dimension four, at a pair production rate of $10^5$ photons per second and the highest lower bound on the fidelity is largest for dimension two, at a pair production rate of $10^5$ photon pairs per second.}
    \label{fig:prod_table_main}
\end{figure*}

Notably, the peak fidelity is consistently observed at dimension two and decreases with increasing dimension. This trend can be attributed to the reconstruction of the density matrix, as it leaves more freedom for off-diagonal entries and can therefore differ more from the maximally entangled target state the higher the dimension is, which is reflected in a smaller lower bound on the fidelity. Within each dimension, the Schmidt number exhibits a non-monotonic behavior, naturally peaking at the same pair production rate as the fidelity. Across different dimensions, however, the maximum achieved Schmidt number is non-decreasing as the dimension increases.

Similar to the key rate per second, the entanglement rate takes on its maximal values at pair production rates significantly different from the ones for the four aforementioned metrics. This discrepancy highlights the fundamental trade-off in QKD: Maximizing the information contained in one coincidence click, i.e. the per-click metrics, most often requires different configurations than maximizing the total information per time unit, which shows that the right optimizaiton is very objective-dependent.

For scenarios with a fixed pair production rate and varying channel loss, all quantities trivially reach their respective maximum at zero loss, while the behavior of the maxima across dimensions remains similar to the case of fixed loss and varying pair production rate, as depicted in Figure \ref{fig:loss_table} in Appendix \ref{app:detailed_results}.

\subsubsection{Discussion of additional optimizations regarding the EoF, fidelity, Schmidt number and key rate per coincidence click}\label{subsubsec:additional_optim}
In Appendix \ref{app:detailed_results} we investigate optimizations regarding further metrics as objective. In the regime with a variable pair production rate, the maximal values obtained when optimizing for different quantities differ only marginally (as shown in Figures \ref{fig:prod2_additional_table} and \ref{fig:prod6_additional_table} in Appendix \ref{app:detailed_results}). In contrast, when optimizing under variable loss, we observe that the maximal values of both the key rate per second and the entanglement rate depend strongly on the quantity with respect to which the bin size is optimized (see Figures \ref{fig:loss2_additional_table} and \ref{fig:loss8_additional_table} in Appendix \ref{app:detailed_results}). This indicates that the choice of the appropriate bin size is of crucial importance in this case.

This observation is further supported by the fact that, within each dimension, the maximum key rate or entanglement rate is not always achieved at zero loss when calculated under the usage of the binsizes optimized for per-coincidence-click-quantities. This stands in contrast to the optimization of the bin size with respect to the key rate per second or the entanglement rate, where all quantities attain their maximum at zero loss. This counterintuitive behavior can presumably be explained by the fact that the optimal bin size is smaller by approximately one order of magnitude when it is optimized for the key rate per second or the entanglement rate, such that for bin sizes in a too high order of magnitude, loss could be beneficial as it could reduce the risk of multi-click events.

Consequently, our additional simulations suggest that it makes little difference for the resulting optimal bin size - especially in higher-dimensional regimes -in regard to which specific per-coincidence-click quantity or to which specific per-second quantity it is optimized, when the goal is to maximize a quantity of the same `type', i.e. per-coincidence-click or per-second (see Figures \ref{fig:prod2_additional_table} to \ref{fig:loss8_additional_table} in Appendix \ref{app:detailed_results}.) However, one has to bear in mind that the best dimension found for a specific quantity might not be the best dimension for a different quantity, not even one of the same type (as indicated by Figures \ref{fig:prod_table} and \ref{fig:loss_table}). At the same time, our additional simulations clearly demonstrate that suboptimal results arise when the bin size is optimized for a per-coincidence-click quantity while aiming to maximize a per-second quantity, or vice versa.

\subsubsection{Discussion of results regarding the Entanglement rate}\label{subsubsec:discuss_entRate}
When optimizing the bin size to maximize the key rate per second, in the regime with variable pair production rate, the simulations within each dimension terminate at pair production rates lower than those at which the entanglement rate would reach its maximum. (For that reason, the optimal pair production rate is not reported in Figure \ref{fig:prod_table}.) Therefore, dedicated simulations are required to properly capture the peak of the entanglement rate. In these additional simulations, where the bin size is optimized with respect to the entanglement rate, the pair production rate is increased until the entanglement rate reaches its maximum and subsequently decreases to nearly zero.

Although it is necessary in the variable-pair-production-rate regime to treat the entanglement rate separately from the key rate per second, the two quantities exhibit very similar behavior in the simulated variable-loss regime. Not only do their respective optimal bin sizes lie within the same order of magnitude, but the final results also differ only slightly depending on whether the bin size is optimized with respect to the key rate per second or the entanglement rate, as shown in Figures \ref{fig:loss2_additional_table} and \ref{fig:loss8_additional_table}. This is probably due to the fact that while for variable pair production rate, the pair production rate at which the key rate per second achieves its maximum is very different from the one where the entanglement rate does, while for variable loss, both quantities attain their maxima at zero loss, which leads to the accompanying optimal bin sizes being similar to each other.

\section{Conclusion}
In this work, we have developed a realistic noise model including temporal jitter, which serves as a critical bridge between theoretical predictions and experimental implementations in high-dimensional QKD by providing two main contributions: First, it enables the a priori identification of optimal experimental configurations before physical implementation. Second, our simulations confirm the empirical observation that in practically relevant regimes, high-dimensional entanglement provides a clear advantage over qubit-based systems, yielding superior results in terms of key rate and entanglement metrics.

Furthermore, to ensure the reliability of entanglement quantification in realistic, noisy experimental scenarios, we have rigorously developed the mathematical framework both for the reconstruction of density matrix elements necessary for lower-bounding the key rate and for the lower-bounding of the EoF and in turn the entanglement rate and of the fidelity and consequently the Schmidt number, ensuring that the results remain physically sound and conservative. Our comprehensive simulations characterize the complex interplay between these quantities and their optimal parameter constellations. These results demonstrate that while higher dimensions offer increased information capacity, their performance is strictly governed by a delicate trade-off between signal-to-noise ratio, timing jitter and the frequency of valid coincidence clicks.
\subsection{Code Availability}
The code used in this work to implement the noise and jitter model and compute the corresponding basis-normalized click matrices will be made publicly available upon acceptance for publication at \url{https://github.com/AlexandraEBergmayr-Mann/noise-model-time-bin-QKD/}.

\begin{acknowledgments}
A. B.-M. would like to thank Florian Kanitschar for providing the code used to calculate the secret key rate from the lower-bounded density matrix elements, based on the analytic method presented in \cite{KanDual}, and for numerous insightful discussions. Furthermore, A. B.-M. is grateful to Dorian Schiffer, Paul Erker, Robert Kindler and Prathwiraj Umesh for fruitful discussions regarding the connection between experiment and theory.

This work has received funding from the Horizon-Europe research and innovation programme under grant agreement No 101070168 (HyperSpace). G.M. acknowledges funding from the Austrian
Research Promotion Agency (FFG) through the
Project NSPT-QKD FO999915265.

\centering
\includegraphics[width=0.25\textwidth]{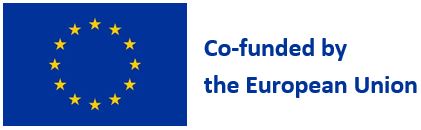} 
\end{acknowledgments}

\bibliography{noise_bib}

\appendix
\onecolumngrid
\section{Definition of the measurement bases}\label{app:bases}
While there are two experimental bases, namely the Time-of-Arrival-basis and the Temporal-Superposition-basis, mathematically these two bases are not mutually orthogonal, such that one has to combine certain elements in order two build mutually orthogonal bases with them.

As detailed in Section IV in \cite{BergKan}, for this purpose we define the basis $\mathcal{B}_0:=\mathcal{B}^A_0 \otimes \mathcal{B}^B_0$ which is formed exactly from the TOA measurement elements,
where 
\begin{equation*}
    \mathcal{B}^{A/B}_0:=\{\ket{i}\}_{i=0}^{d-1}
\end{equation*} span the local time-bin subspaces. Another theoretical basis, $\mathcal{B}_1:=\mathcal{B}^A_1 \otimes \mathcal{B}^B_1$, is built by combining only TSUP measurement elements,
\begin{equation*}
    \mathcal{B}^{A/B}_1:=\left\{\frac{\ket{(2k-1)}\pm \ket{2(k-1)}}{\sqrt{2}}\right\}_{k=1}^{\frac{d}{2}},
\end{equation*}
while a third one can be found by considering both TOA and TSUP measurements, 
\begin{equation*}
    \mathcal{B}^{A/B}_2:= \left\{\frac{\ket{2k}\pm \ket{2k-1}}{\sqrt{2}}\right\}_{k=1}^{\frac{d}{2}-1}\cup \{\ket{0}, \ket{d-1}\}.
\end{equation*}

\section{Simulation of the click matrices}\label{app:click_mat_sim}
Starting from the expected pre-jitter state $\rho_1$ (as modeled in Section \ref{sec:calcNoiseLossProbs}), we simulate the coincidence click matrices expected in the absence of timing jitter. Following the methodology in \cite{BergKan} (equ. 17 - 25), we get the following calculation of the TOA-TOA- click matrix $cc$ (meaning that both Alice and Bob measure in the TOA-basis), the TSUP-TSUP- click matrix $dd_{ij}$ where detectors $i, j \in \{1,2\}$ click on Alice's resp. Bob's side, the TOA-TSUP- click matrix $cd_j$, $j \in \{1,2\}$, where detector $j$ clicks on Bob's side when measuring in the TSUP-basis while Alice's measurement being done in the TOA-basis, and the TSUP-TOA- click matrix $dc_i$, $i \in \{1,2\}$, where detector $i$ clicks on Alice's side.
\begin{align*}
    cc(m+1,n+1) := \text{Tr}(\rho_1\ketbra{m+1,n+1}) = \rho_1(m d + n + 1, m d + n + 1), \text{ for } m, n = 0, \dotsc, d-1
\end{align*}
\begin{align*}
    dd_{11}(m+1,n+1) &:= \frac{1}{4}\text{Tr}(\rho_1\ketbra{(m+1)+,(n+1)+})\\ 
    &= \frac{1}{16}[\rho_1(md+n+1, md+n+1) + \rho_1(md+n+1, md+n) + \rho_1(md+n+1, (m-1)d+n+1) \\ &+ \rho_1(md+n+1, (m-1)d+n) + \rho_1(md+n, md+n+1) + \rho_1(md+n, md+n) \\ &+ \rho_1(md+n, (m-1)d+n+1) + \rho_1(md+n, (m-1)d+n)
    + \rho_1((m-1)d+n+1, md+n+1)\\ &+ \rho_1((m-1)d+n+1, md+n) + \rho_1((m-1)d+n+1, (m-1)d+n+1)\\ &+ \rho_1((m-1)d+n+1, (m-1)d+n) + \rho_1((m-1)d+n, md+n+1) + \rho_1((m-1)d+n, md+n) \\&+ \rho_1((m-1)d+n, (m-1)d+n+1) + \rho_1((m-1)d+n, (m-1)d+n)], \text{ for } m,n = 1, \dotsc, d-1 \\
    dd_{12}(m+1,n+1) &:= \frac{1}{4}\text{Tr}(\rho_1\ketbra{(m+1)+,(n+1)-})\\
    &= \frac{1}{16}[\rho_1(md+n+1, md+n+1) - \rho_1(md+n+1, md+n) + \rho_1(md+n+1, (m-1)d+n+1)\\ &- \rho_1(md+n+1, (m-1)d+n) - \rho_1(md+n, md+n+1) + \rho_1(md+n, md+n)\\ &- \rho_1(md+n, (m-1)d+n+1) + \rho_1(md+n, (m-1)d+n) + \rho_1((m-1)d+n+1, md+n+1)\\ &- \rho_1((m-1)d+n+1, md+n) + \rho_1((m-1)d+n+1, (m-1)d+n+1)\\ &- \rho_1((m-1)d+n+1, (m-1)d+n)- \rho_1((m-1)d+n, md+n+1) + \rho_1((m-1)d+n, md+n)\\ &- \rho_1((m-1)d+n, (m-1)d+n+1)  + \rho_1((m-1)d+n, (m-1)d+n)], \text{ for } m,n = 1, \dotsc, d-1\\
    dd_{21}(m+1,n+1) &:= \frac{1}{4}\text{Tr}(\rho_1\ketbra{(m+1)-,(n+1)+})\\
    &= \frac{1}{16}  [\rho_1(md+n+1, md+n+1) + \rho_1(md+n+1, md+n) - \rho_1(md+n+1, (m-1)d+n+1)\\ &- \rho_1(md+n+1, (m-1)d+n) + \rho_1(md+n, md+n+1) + \rho_1(md+n, md+n)\\ &- \rho_1(md+n, (m-1)d+n+1) - \rho_1(md+n, (m-1)d+n) - \rho_1((m-1)d+n+1, md+n+1)\\ &- \rho_1((m-1)d+n+1, md+n) + \rho_1((m-1)d+n+1, (m-1)d+n+1)\\ &+ \rho_1((m-1)d+n+1, (m-1)d+n) - \rho_1((m-1)d+n, md+n+1) - \rho_1((m-1)d+n, md+n)\\ &+ \rho_1((m-1)d+n, (m-1)d+n+1)  + \rho_1((m-1)d+n, (m-1)d+n)], \text{ for } m,n = 1, \dotsc, d-1\\
    dd_{22}(m+1,n+1) &:= \frac{1}{4}\text{Tr}(\rho_1\ketbra{(m+1)-,(n+1)-})\\
    &= \frac{1}{16}[\rho_1(md+n+1, md+n+1) - \rho_1(md+n+1, md+n) - \rho_1(md+n+1, (m-1)d+n+1)\\ &+ \rho_1(md+n+1, (m-1)d+n) - \rho_1(md+n, md+n+1) + \rho_1(md+n, md+n)\\ &+ \rho_1(md+n, (m-1)d+n+1) - \rho_1(md+n, (m-1)d+n) - \rho_1((m-1)d+n+1, md+n+1)\\ &+ \rho_1((m-1)d+n+1, md+n) + \rho_1((m-1)d+n+1, (m-1)d+n+1)\\ &- \rho_1((m-1)d+n+1, (m-1)d+n) + \rho_1((m-1)d+n, md+n+1) - \rho_1((m-1)d+n, md+n)\\ &- \rho_1((m-1)d+n, (m-1)d+n+1)  + \rho_1((m-1)d+n, (m-1)d+n)], \text{ for } m,n = 1, \dotsc, d-1
\end{align*}
\begin{align*}
    cd_{1}(m+1,n+1) &:= \frac{1}{2}\text{Tr}(\rho_1\ketbra{(m+1),(n+1)+})\\
    &= \frac{1}{4} [\rho_1(md+n+1, md+n+1) + \rho_1(md+n, md+n+1) + \rho_1(md+n+1, md+n)\\ &+ \rho_1(md+n, md+n)] \text{ for } m = 0, \dotsc d-1, n = 1, \dotsc d-1\\
    cd_{2}(m+1,n+1) &:= \frac{1}{2}\text{Tr}(\rho_1\ketbra{(m+1),(n+1)-})\\
    &= \frac{1}{4} [\rho_1(md+n+1, md+n+1) - \rho_1(md+n, md+n+1) - \rho_1(md+n+1, md+n)\\ &+ \rho_1(md+n, md+n)] \text{ for } m = 0, \dotsc d-1, n = 1, \dotsc d-1
\end{align*}
\begin{align*}
    dc_{1}(m+1,n+1) &:= \frac{1}{2}\text{Tr}(\rho_1\ketbra{(m+1)+,(n+1)})\\
    &= \frac{1}{4} [\rho_1(md+n+1, md+n+1) + \rho_1((m-1)d+n+1, md+n+1)\\ &+ \rho_1(md+n+1, (m-1)d+n+1) + \rho_1((m-1)d+n+1, (m-1)d+n+1)]\\ &\text{ for } m = 1, \dotsc, d-1, n = 0, \dotsc, d-1\\
    dc_{2}(m+1,n+1) &:= \frac{1}{2}\text{Tr}(\rho_1\ketbra{(m+1)-,(n+1)})\\
    &= \frac{1}{4} [\rho_1(md+n+1, md+n+1) - \rho_1((m-1)d+n+1, md+n+1)\\ &- \rho_1(md+n+1, (m-1)d+n+1) + \rho_1((m-1)d+n+1, (m-1)d+n+1)]\\ &\text{ for } m = 1, \dotsc, d-1, n = 0, \dotsc, d-1
\end{align*}

It should be noted that while all elements are defined for the $cc$ matrix, the other click matrices lack the definition of certain elements due to the non-existence of what would be the elements $\ket{0+}, \ket{0-}$ (in the case of the usage of a pulsed laser, where no preceding time-bin exists to form a superposition with the first bin, which we are assuming here). Consequently, these missing entries are either set to zero or omitted within their respective matrices. Since these elements are neither required nor utilized in the subsequent analysis, their absence does not impact the integrity of the model.

Furthermore, one has to be aware that these coincidence click matrices are not directly associated with mathematical bases, but that some of their entries belong to (the multiple of) elements of (one or more of) three distinct bases, each spanning the whole $d^2$-dimensional Hilbert space of their shared temporal quantum state (as detailed in Section IV for \cite{BergKan}). Therefore, the entries of the click matrices above must `normalized' according to which basis they belong to. In our framework, this normalization has to be done after the jitter has been considered, yielding `basis-normalized' coincidence click matrices $\widetilde{cc}(i,j), \widetilde{dd}_{a,b}(i,j), \widetilde{cd}_b(i,j), \widetilde{dc}_a(i,j), \, a,b \in {1,2}$. Among these, only $\widetilde{cc}$ and $\widetilde{dd}_{a,b}(i,j)$ are required for the reconstruction of certain (off-)diagonal elements of the (real part of) the density matrix.

\section{Details on the normalization factors and basis-normalized click matrices}
\label{app:normalization}
Here we present the details on the factors $\mathrm{CM0}$, $\mathrm{CM1}$ and $\mathrm{CM2}$, which the jittered coincidence clicks are divided by in order to obtain their `basis-normalized' counterparts. We follow the methodology in \cite{BergKan}, Section IV, where equations 27, 29 and 31 yield the normalization factors.

Each of the factors is calculated by summing over the elements of one of the matrices $\mathcal{M}_0, \mathcal{M}_1$ and $\mathcal{M}_2$, which are built out of all the jittered coincidence clicks belonging to one of the three bases respectively.

For basis $\mathcal{B}_0$, this matrix reads
\begin{equation*}
    \mathcal{M}_0(i,j) = cc^J(i,j), \qquad i,j = 0, \dotsc, d-1,
\end{equation*}
for $\mathcal{B}_1$ we have
\begin{equation*}
    \mathcal{M}_1(i,j) = 4 \cdot 
\begin{cases} 
dd^J_{1,1}(i+1, j+1), & \text{if } i \text{ even}, j \text{ even}, \\
dd^J_{1,2}(i+1, j),   & \text{if } i \text{ even}, j \text{ odd}, \\
dd^J_{2,1}(i, j+1),   & \text{if } i \text{ odd}, j \text{ even}, \\
dd^J_{2,2}(i, j),     & \text{if } i \text{ odd}, j \text{ odd},
\end{cases}
\end{equation*}
where $i,j = 0, \dotsc, d-1$. Finally, the matrix built out of clicks corresponding to basis $\mathcal{B}_2$ is
\begin{equation*}
    \mathcal{M}_2(i,j) =    \begin{cases} 
        cc^J(i,j), & \text{if } i,j \in \{0,d-1\}\\
        2 cd^J_1(i,j+1), & \text{if } i\in \{0,d-1\} \text{ and } 1\leq j\leq d-3 \text{ and } j \text{ odd}, \\
        2 cd^J_2(i,j), & \text{if } i\in \{0,d-1\} \text{ and } 1\leq j\leq d-2 \text{ and } j \text{ even}, \\
        2 dc^J_1(i+1,j), & \text{if } j\in \{0,d-1\} \text{ and } 1\leq i\leq d-3 \text{ and } i \text{ odd}, \\
        2 dc^J_2(i,j), & \text{if } j\in \{0,d-1\} \text{ and } 1\leq i\leq d-2 \text{ and } i \text{ even} \\
        4 dd^J_{1,1}(i+1,j+1), & \text{if } 1\leq i,j\leq d-2 \text{ and } i,j \text{ odd}, \\
        4 dd^J_{1,2}(i+1,j), & \text{if } 1\leq i,j\leq d-2 \text{ and } i \text{ odd}, j \text{ even}, \\
        4 dd^J_{2,1}(i,j+1), & \text{if } 1\leq i,j\leq d-2 \text{ and } i \text{ even}, j \text{ odd}, \\
        4 dd^J_{2,2}(i,j), & \text{if } 1\leq i,j\leq d-2 \text{ and } i, j \text{ even}.
    \end{cases}
\end{equation*}

Next, the respective sums over these three matrices are built, yielding the factor $\mathrm{CM0}$ as the sum over all elements of $\mathcal{M}_0$ and similarly the factors $\mathrm{CM1}$ and $\mathrm{CM2}$. Lastly, these factors are used to get the `basis-normalized' coincidence click matrices, 
\begin{align*}
    \widetilde{cc}(i,j)&:=\frac{cc^J}{\mathrm{CM0}}, \text{ for } 0 \leq i,j \leq d-1,\\ 
    \widetilde{dd}_{k,l}(i,j)&:=4\begin{cases}
        \frac{dd^J_{k,l}(i,j)}{\mathrm{CM1}}, \text{ for } 1 \leq i,j \leq d-1 \text{ and } i,j \text{ odd},\\
        \frac{dd^J_{k,l}(i,j)}{\mathrm{CM2}}, \text{ for } 2 \leq i,j \leq d-2 \text{ and } i,j \text{ even},
    \end{cases} k,l \in \{1,2\}\\
    \widetilde{cd}_k(i,j) &:= 2\frac{cd^J_k(i,j)}{\mathrm{CM2}}, \text{ for } i \in \{0,d-1\} \text{ and } 2 \leq j \leq d-2 \text{ and } j \text{ even and } k \in \{1,2\}\\
    \widetilde{dc}_k(i,j) &:= 2\frac{dc^J_k(i,j)}{\mathrm{CM2}}, \text{ for } j \in \{0,d-1\} \text{ and } 2 \leq i \leq d-2 \text{ and } i \text{ even and } k \in \{1,2\}
\end{align*}

Note that the matrices are not fully populated, which is of no concern, since they contain all the elements needed to reconstruct the (relevant parts of the) density matrix.

Furthermore, note that the belonging of an element to more than one of the three matrices $\mathcal{M}_i, \, i \in \{1,2,3\}$ in principle yields more than one option for its corresponding normalization factor to choose. In the simulations, this is of no concern since $\theta$, the probability of directing a local photon to a TOA measurement, which would yield a imbalanced weighting of the absolute TOA- and TSUP-click-numbers in practice, is involved neither in the derivation of the coincidence click matrices nor in their basis-normalized counterparts, but are taken into account only later for the transition from per-click-quantities to per-frame-quantities. However, when experimental data providing coincidence click matrices with absolute numbers corresponding to a setup for which $\theta \neq 0.5$ is involved, this imbalance has to be accounted for before the calculation of the normalization factors $\mathrm{CM0}$, $\mathrm{CM1}$ and $\mathrm{CM2}$.

Lastly, it should be noted that the $\widetilde{cd}-$ and $\widetilde{cd}-$ matrices are given here only for completeness, but are not necessary for our subsequent reconstruction of the density matrix, which makes use only of the $\widetilde{cc}-$ and $\widetilde{dd}-$ matrices.

\section{Derivation of the coincidence click probabilities}\label{app:prob_calc}
Here, we derive the formulas for the probabilities defined in Section \ref{sec:calcNoiseLossProbs}, which are used to predict $\rho_1$, the density matrix of the state that we expect to reach the detectors. The parameters used were introduced in Sections \ref{subsec:tunableParams} and \ref{subsec:noise_jitter}.
\begin{equation}\label{source_source}
    \begin{split}
    \begin{aligned}
    p(A\varphi, B\varphi) &= \left[ \sum\limits_{n=1}^{\infty} P_{\text{prod}}(n) \sum\limits_{j=1}^n \binom{n}{j}(P_{\text{surv}}^A)^j (P_{\text{surv}}^B)^j \binom{j}{1} \eta_A^1 \eta_B^1 (1-\eta_A)^{j-1} (1-\eta_B)^{j-1} \right. \\ 
    &\left.\times \sum\limits_{k=0}^{n-j}\binom{n-j}{k}(P_{\text{surv}}^A)^k (1-P_{\text{surv}}^A)^{n-j-k}(1-\eta_A)^k\sum\limits_{l=0}^{n-j-k} \binom{n-j-k}{l}(P_{\text{surv}}^B)^l (1-P_{\text{surv}}^B)^{n-j-l}   (1-\eta_B)^l\right]\\
    & \times \left[\sum\limits_{m=0}^{\infty} \sum\limits_{j=0}^{\infty}P_{\text{env}}^A(m)P_{\text{env}}^B(j) \binom{m}{0} \binom{j}{0}\eta_A^0 \eta_B^0 (1-\eta_A)^m (1-\eta_B)^j\right]P_{\text{dark}}^A(0)P_{\text{dark}}^B(0)\\
    &= \dotsc = \left[\sum\limits_{n=1}^{\infty} P_{\text{prod}}(n)\cdot n\, \cdot P_{\text{surv}}^A \eta_A (1-P_{\text{surv}}^A \eta_A)^{n-1} P_{\text{surv}}^B \eta_B (1-P_{\text{surv}}^B \eta_B)^{n-1}\right]\\
    & \times \left[\sum\limits_{m=0}^{\infty} \sum\limits_{j=0}^{\infty}P_{\text{env}}^A(m)P_{\text{env}}^B(j)(1-\eta_A)^m (1-\eta_B)^j\right]P_{\text{dark}}^A(0)P_{\text{dark}}^B(0)
    \end{aligned} 
    \end{split}
\end{equation}

The sum $\sum\limits_{n=1}^{\infty}$ together with $P_{\text{prod}}(n)$ consider that at least one source photon pair has to be produced in case a source photon gets detected and goes up to infinity, because in principle, there's no upper bound on the number of source photons that could be produced per time-frame (though in practice, of course $P_{prod}(n)$ will vanish for large $n$). The index $j$ runs over all possible numbers of source photon pairs that can have survived pairwise. There are $\binom{n}{j}$ possibilities to choose $j$ surviving photon pairs out of the $n$ produced ones. $(P_{\text{surv}^A})^j$ and $(P_{\text{surv}^B})^j$ are the probabilities for $j$ photons to survive on Alice's resp. Bob's side. It is also possible that some of the $n-j$ remaining photon pairs are transmitted only half, i.e. that not the whole pair is transmitted, but only a single photon on Alice's or Bob's side. The index $k$ runs over all the possible numbers of such additional photons to survive on Alice's side. There are $\binom{n-j}{k}$ possibilities to choose $k$ such single-photons out of the $n-j$ source photons which are not transmitted as a whole pair. $(P_{\text{surv}}^A)^k$ is the probability for $k$ such photons to survive Alice's channel and $(1-P_{\text{surv}}^A)^{n-j-k}$ is the probability that all the other $n-j-k$ photons which were sent to Alice by the source got lost on their way. Additional photons which are the components of a pair that isn't transmitted to both Alice and Bob also have to be taken into account for Bob. For him, these photons have to be chosen out of the set of the $n$ source photon pairs excluding the $j$ photons that are transmitted as a whole pair and excluding the $k$ pairs a component of which reaches only Alice. Therefore, there are $\binom{n-j-k}{l}$ possibilities to choose $l$ additional photons that make their way only to Bob. $(P_{\text{surv}}^B)^l$ is the probability for the $l$ photons to survive and $(1-P_{\text{surv}}^B)^{n-j-l}$ is the probability that all of the other photons that were sent out to Bob get lost.
Next, $\binom{j}{1}$, the number of options to choose one photon pair that gets detected out of the $j$ successfully transmitted ones, is accounted for. This number is multiplied by $\eta_A^1$ and $\eta_B^1$, the probability for Alice's resp. Bob's detector to click in the presence of a photon. $(1-\eta_A)^{j-1}$ and $(1-\eta_B)^{j-1}$ are the probabilities for all the other $j-1$ to be detected neither on Alice's nor on Bob's side. (Note that if only one of the photons were detected and no click stemming from other sources would happen on the other side, we wouldn't have a coincidence click, so we don't have to consider this case. The other possibilities - that only one photon of the pair is detected and the detector on the other side yields a click due to a photon from a different pair or due to other causes - are covered by subsequent probabilities below.) The term in the second to last line of the equation is necessary to assure that no additional (single- or coincidence-) click is caused by any other possible source. Finally, the last two lines are just a simplification of the formula we just discussed.

Next, we elaborate how the probability $p(A\varphi, B\tilde{\varphi})$ can be calculated. There are four possible subcases which contribute to this case: Firstly, Alice and Bob might each detect a photon stemming from successfully transmitted photon pairs, but the detected photons originate from different pairs. Secondly, Alice might detect a photon stemming from a successfully transmitted photon pair while Bob detects one from a pair where only his component survived. Thirdly, the second case, with the roles of Alice and Bob interchanged, might happen. Lastly, both parties might detect photons from pairs that survived only for their respective side. These considerations lead to the following formula.
\begin{align*} 
        &p(A\varphi, B\tilde{\varphi}) = \left[\sum_{n=2}^{\infty}P_{\text{prod}}(n) \left[\sum_{j=2}^n\binom{n}{j}(P^A_{\text{surv}})^j (P^B_{\text{surv}})^j \binom{j}{1} \binom{j-1}{1}\eta^1_A \eta^1_B (1-\eta_A)^{j-1}(1-\eta_B)^{j-1}\right.\right.
        \\
        &\left.\left.\times\sum_{k=0}^{n-j}\binom{n-j}{k}(P^A_{\text{surv}})^k(1-P^A_{\text{surv}})^{n-j-k} \binom{k}{0} \eta_A^0 (1-\eta_A)^k \sum\limits_{l=0}^{n-j-k}\binom{n-j-k}{l}(P^B_{\text{surv}})^l(1-P^B_{\text{surv}})^{n-j-l} \binom{l}{0} \eta_B^0 (1-\eta_B)^l\right.\right.
        \\
        &\left.\left.+\sum\limits_{j=1}^{n-1}\binom{n}{j}(P_{\text{surv}}^A)^j (P_{\text{surv}}^B)^j \binom{j}{1} \eta_A^1 \eta_B^0 (1-\eta_A)^{j-1} (1-\eta_B)^j \sum_{k=0}^{n-j-1} \binom{n-j}{k}(P_{\text{surv}}^A)^k (1-P_{\text{surv}}^A)^{n-j-k}\binom{k}{0}\eta_A^0 (1-\eta_A)^k \right.\right.\\
        &\left.\left.\times \sum\limits_{l=1}^{n-j-k} \binom{n-j-k}{l} (P_{\text{surv}}^B)^l (1-P_{\text{surv}}^B)^{n-j-l}  \binom{l}{1}\eta_B^1 (1-\eta_B)^{l-1} \right.\right.\\
        &\left.\left.+\sum\limits_{j=1}^{n-1}\binom{n}{j}(P_{\text{surv}}^A)^j (P_{\text{surv}}^B)^j \binom{j}{1} \eta_A^0 \eta_B^1 (1-\eta_A)^j (1-\eta_B)^{j-1}\sum\limits_{k=1}^{n-j}\binom{n-j}{k}(P_{\text{surv}}^A)^k (1-P_{\text{surv}}^A)^{n-j-k}\binom{k}{1}\eta_A^1 (1-\eta_A)^{k-1}\right.\right.\\
        &\left.\left.\times \sum\limits_{l=0}^{n-j-k}\binom{n-j-k}{l}(P_{\text{surv}}^B)^l (1-P_{\text{surv}}^B)^{n-j-l}\binom{l}{0}\eta_B^0 (1-\eta_B)^l \right.\right.\\
        &\left.\left.+\sum\limits_{j=0}^{n-2}\binom{n}{j}(P^A_{\text{surv}})^j(P^B_{\text{surv}})^j \binom{j}{0}\eta^0_A\eta^0_B (1-\eta_A)^j(1-\eta_B)^j\sum\limits_{k=1}^{n-j-1}\binom{n-j}{k}(P^A_{\text{surv}})^k(1-P^A_{\text{surv}})^{n-j-k}\binom{k}{1}\eta_A^1(1-\eta_A)^{k-1}\right.\right.
        \\
        &\left.\left. \times\sum\limits_{l=1}^{n-j-k}\binom{n-j-k}{l}(P^B_{\text{surv}})^l(1-P^B_{\text{surv}})^{n-j-l}\binom{l}{1}\eta_B^1(1-\eta_B)^{l-1} \right]\right]\\
        & \times \left[\sum\limits_{m=0}^{\infty} \sum\limits_{j=0}^{\infty}P_{\text{env}}^A(m)P_{\text{env}}^B(j) \binom{m}{0} \binom{j}{0}\eta_A^0 \eta_B^0 (1-\eta_A)^m (1-\eta_B)^j\right]P_{\text{dark}}^A(0)P_{\text{dark}}^B(0)
        \\ 
        &= \dotsc = \left[\sum\limits_{n=2}^{\infty}P_{\text{prod}}(n)n(n-1)P_{\text{surv}}^A\eta_A(1-P_{\text{surv}}^A\eta_A)^{n-1}P_{\text{surv}}^B\eta_B(1-P_{\text{surv}}^B\eta_B)^{n-1}\right]\\
        & \times \left[\sum\limits_{m=0}^{\infty} \sum\limits_{j=0}^{\infty}P_{\text{env}}^A(m)P_{\text{env}}^B(j) (1-\eta_A)^m (1-\eta_B)^j\right]P_{\text{dark}}^A(0)P_{\text{dark}}^B(0)
\end{align*}

Here, the first two lines account for the first case mentioned above,
the third and fourth line cover the second scenario, the fifth and sixth line correspond to the third case and the seventh to ninth line cover the fourth case discussed, where the last line before the second equation sign ensures that no environmental photons or dark counts contribute a click. The formula that follows afterwards in the last two lines is just a simplification of the upper equation.

Since this calculation inherently covers the exchange of $\varphi$ and $\tilde{\varphi}$, no separate treatment for $p(A\tilde{\varphi}, B\varphi)$ is required.

The analysis of the probabilities for the remaining events follows the same pattern, but is more straightforward, so we only give the results and omit a detailed explanation for brevity here.

\begin{align*}
    p(A\varphi, B\delta) &= \left[\sum\limits_{n=1}^{\infty}\sum\limits_{k=1}^n \sum\limits_{l=0}^n P_{\text{prod}}(n) \binom{n}{k} \binom{n}{l}(P_{\text{surv}}^A)^k (P_{\text{surv}}^B)^l (1-P_{\text{surv}}^A)^{n-k} (1-P_{\text{surv}}^B)^{n-l} \right. \\ 
    & \qquad \qquad \qquad \qquad \quad \; \times \left.\binom{k}{1} \binom{l}{0}\eta_A^1 \eta_B^0 (1-\eta_A)^{k-1}(1-\eta_B)^l  \right]\\ 
    &\times  \left[\sum\limits_{m=0}^{\infty} \sum\limits_{j=0}^{\infty} P_{\text{env}}^A(m) P_{\text{env}}^B(j) \binom{m}{0} \binom{j}{0}\eta_A^0 \eta_B^0 (1-\eta_A)^m (1-\eta_B)^j \right] P_{\text{dark}}^A(0) P_{\text{dark}}^B(1)\\
    &= \cdots = \left[\sum\limits_{n=1}^{\infty}P_{\text{prod}}(n) n\, P_{\text{surv}}^A \eta_A (1-P_{\text{surv}}^A \eta_A)^{n-1} (1-P_{\text{surv}}^B \eta_B)^n\right]\\
    &\times \left[\sum\limits_{m=0}^{\infty} \sum\limits_{j=0}^{\infty} P_{\text{env}}^A(m) P_{\text{env}}^B(j) (1-\eta_A)^m (1-\eta_B)^j\right]P_{\text{dark}}^A(0) P_{\text{dark}}^B(1)
\end{align*}

\begin{align*}
p(A\delta, B\varphi): \text{ looks the same as $P(A\varphi, B\delta)$, but with the roles of $A$ and $B$ interchanged} 
\end{align*}

\begin{align*}
    p(A\varphi, B\xi) &= \left[\sum\limits_{n=1}^{\infty}\sum\limits_{k=1}^n \sum\limits_{l=0}^n P_{\text{prod}}(n)\binom{n}{k} \binom{n}{l} (P_{\text{surv}}^A)^k (P_{\text{surv}}^B)^l (1-P_{\text{surv}}^A)^{n-k} (1-P_{\text{surv}}^B)^{n-l} \right. \\ 
    &\qquad \qquad \qquad \qquad \qquad \times \left.\binom{k}{1} \binom{l}{0}\eta_A^1 \eta_B^0 (1-\eta_A)^{k-1}(1-\eta_B)^l  \right]\\ 
    &\times  \left[\sum\limits_{m=0}^{\infty} \sum\limits_{j=1}^{\infty} P_{\text{env}}^A(m) P_{\text{env}}^B(j)\binom{m}{0} \binom{j}{1} \eta_A^0 (1-\eta_A)^m \eta_B^1 (1-\eta_B)^{j-1} \right] P_{\text{dark}}^A(0) P_{\text{dark}}^B(0)\\
    & = \dotsc = \left[\sum\limits_{n=1}^{\infty} P_{\text{prod}}(n) n\, P_{\text{surv}}^A \eta_A (1-P_{\text{surv}}^A \eta_A)^{n-1} (1-P_{\text{surv}}^B \eta_B)^n\right] \\
    & \times \left[\sum\limits_{m=0}^{\infty} \sum\limits_{j=1}^{\infty} P_{\text{env}}^A(m) P_{\text{env}}^B(j) \,j\,(1-\eta_A)^m \eta_B  (1-\eta_B)^{j-1}\right] P_{\text{dark}}^A(0) P_{\text{dark}}^B(0)
\end{align*}

\begin{align*}
p(A\xi, B\varphi): \text{ looks the same as $P(A\varphi, B\xi)$, but with the roles of $A$ and $B$ interchanged} 
\end{align*}

\begin{align*}
        p(A\delta, B\xi) &= \left[\sum\limits_{n=0}^{\infty}\sum\limits_{k=0}^n\sum\limits_{l=0}^n P_{\text{prod}}(n)\binom{n}{k} \binom{n}{l} (P_{\text{surv}}^A)^k (P_{\text{surv}}^B)^l (1-P_{\text{surv}}^A)^{n-k} (1-P_{\text{surv}}^B)^{n-l} \binom{k}{0} \binom{l}{0}\eta_A^0 \eta_B^0 \right. \\ & \qquad \qquad \qquad \qquad \qquad \times\left.(1-\eta_A)^k(1-\eta_B)^l \right]\\
        &\times \left[\sum\limits_{m=0}^{\infty}\sum\limits_{j=1}^{\infty} P_{\text{env}}^A(m) P_{\text{env}}^B(j) \binom{m}{0} \binom{j}{1} \eta_A^0 \eta_B^1 (1-\eta_A)^m (1-\eta_B)^{j-1} \right] P_{\text{dark}}^A(1) P_{\text{dark}}^B(0)\\
        & = \dotsc = \left[\sum\limits_{n=0}^{\infty}P_{\text{prod}}(n)(1-P_{\text{surv}}^A \eta_A)^n (1-P_{\text{surv}}^B \eta_B)^n\right] \\
        &\times \left[\sum\limits_{m=0}^{\infty}\sum\limits_{j=1}^{\infty}  P_{\text{env}}^A(m) P_{\text{env}}^B(j) \,j \,(1-\eta_A)^m \eta_B  (1-\eta_B)^{j-1}\right] P_{\text{dark}}^A(1) P_{\text{dark}}^B(0)
\end{align*}

\begin{align*}
p(A\xi, B\delta): \text{ looks the same as $p(A\delta, B\xi)$, but with the roles of $A$ and $B$ interchanged} 
\end{align*}

\begin{align*}
        p(A\xi, B\xi) &= \left[\sum\limits_{n=0}^{\infty}\sum\limits_{k=0}^n\sum\limits_{l=0}^n P_{\text{prod}}(n) \binom{n}{k} \binom{n}{l}(P_{\text{surv}}^A)^k (P_{\text{surv}}^B)^l (1-P_{\text{surv}}^A)^{n-k} (1-P_{\text{surv}}^B)^{n-l} \binom{k}{0} \binom{l}{0}\eta_A^0 \eta_B^0 \right. \\ & \qquad \qquad \qquad \qquad \qquad \times\left.(1-\eta_A)^k(1-\eta_B)^l \right]\\
        &\times \left[\sum\limits_{m=1}^{\infty}\sum\limits_{j=1}^{\infty} P_{\text{env}}^A(m) P_{\text{env}}^B(j)\binom{m}{1} \binom{j}{1} \eta_A^1 (1-\eta_A)^{m-1} \eta_B^1 (1-\eta_B)^{j-1} \right] P_{\text{dark}}^A(0) P_{\text{dark}}^B(0)\\
        & = \dotsc = \left[\sum\limits_{n=0}^{\infty} P_{\text{prod}}(n)(1-P_{\text{surv}}^A \eta_A)^n (1-P_{\text{surv}}^B \eta_B)^n\right] \\ & \times \left[\sum\limits_{m=1}^{\infty}\sum\limits_{j=1}^{\infty} P_{\text{env}}^A(m) P_{\text{env}}^B(j) \,m \,j\, \eta_A (1-\eta_A)^{m-1}  \eta_B(1-\eta_B)^{j-1} \right] P_{\text{dark}}^A(0) P_{\text{dark}}^B(0)
\end{align*}

\begin{align*}
        p(A\delta, B\delta) &= \left[\sum\limits_{n=0}^{\infty}\sum\limits_{k=0}^n\sum\limits_{l=0}^n P_{\text{prod}}(n) \binom{n}{k} \binom{n}{l}(P_{\text{surv}}^A)^k (P_{\text{surv}}^B)^l (1-P_{\text{surv}}^A)^{n-k} (1-P_{\text{surv}}^B)^{n-l} \binom{k}{0} \binom{l}{0}\eta_A^0 \eta_B^0\right. \\ & \qquad \qquad \qquad \qquad \qquad \times\left.(1-\eta_A)^k(1-\eta_B)^l \right]\\
        &\times \left[\sum\limits_{m=0}^{\infty}\sum\limits_{j=0}^{\infty} P_{\text{env}}^A(m) P_{\text{env}}^B(j) \binom{m}{0} \binom{j}{0}\eta_A^0 \eta_B^0 (1-\eta_A)^m (1-\eta_B)^{j} \right] P_{\text{dark}}^A(1) P_{\text{dark}}^B(1)\\
        &= \dotsc = \left[\sum\limits_{n=0}^{\infty} P_{\text{prod}}(n)(1-P_{\text{surv}}^A \eta_A)^n (1-P_{\text{surv}}^B \eta_B)^n\right]\\
        &\times \left[\sum\limits_{m=0}^{\infty}\sum\limits_{j=0}^{\infty} P_{\text{env}}^A(m) P_{\text{env}}^B(j) (1-\eta_A)^m (1-\eta_B)^{j}  \right] P_{\text{dark}}^A(1) P_{\text{dark}}^B(1)
\end{align*}

\section{Calculation of the error correction term \texorpdfstring{$H(A|B)_{\text{cc}}$}{H(A|B)cc}}\label{app:ecterm}
As established in Section \ref{sec:lower_bound_key}, the error correction term is derived directly from the simulated (basis-normalized) jittered TOA-TOA-click matrix (derived in Section \ref{sec:jitter_simulation}), which is used for the reconstruction of the density matrix later. The conditional entropy is defined as \begin{equation}
    H(X|Y):=-\sum\limits_{x,y}p(x,y)\text{log}_2\left(\frac{p(x,y)}{p(y)}\right). 
\end{equation}

To compute $H(A|B)_{\text{cc}}$, the jittered click matrix is treated as the joint probability distribution of Alice’s and Bob’s measurements, with $x$ and $y$ above denoting the time-bin indices $a, b \in \{0, \dotsc, d-1\}$. The marginal distribution $p(b)$, representing the probability of Bob detecting a click in time-bin $b$, is obtained by summing over the rows of the $b+1$st column.

For the numerical implementation, we ensure that the evaluation remains stable in the presence of zero-probability events by omitting the terms for which we have $p(a,b) = 0$, in consistency with the limit $\underset{p\rightarrow 0}{\text{lim}}\left(p\cdot \text{log}_2(p)\right) \rightarrow 0$.

This ensures that the calculation of $H(A|B)_{\text{cc}}$ correctly reflects the uncertainty Bob has about Alice's bit string, which is the fundamental quantity for determining the cost of error correction in the QKD protocol.

\section{Conversion of key rate per coincidence click to key rate per second and from EoF to Entanglement rate}
\label{app:keyrate_per_second}
To evaluate the performance of practical QKD implementations across various dimensions $d$, it is necessary to compare the secret key rate per second, $r_{\text{sec}}$, rather than the key rate per coincidence click, $r_{\text{cc}}$, because typically it will be the case that the key rate per coincidence click is larger for higher dimensions, while the key rate per second is better for lower dimensions, since a larger dimension inherently requires longer time-frames, thereby reducing the number of frames that can be transmitted per second.

As can be easily derived, the formula for the conversion is given by
\begin{equation}
    r_{\text{sec}} = r_{\text{frame}} n_{\text{frames}},
\end{equation}
where $r_{\text{sec}}$ denotes the key rate per second, $n_{\text{frames}}$ denotes the number of frames per second (when including one additional paused bin between frames) and $r_{\text{frame}}$ denotes the key rate per frame, given by
\begin{equation}
    r_{\text{frame}} = r_{\text{cc}} P_{\text{click}} p_{\text{CC}},
\end{equation}
where $r_{\text{cc}}$ stands for the key rate per coincidence click, $P_{\text{click}}$ for the probability that exactly one coincidence click occurs within a frame and $p_{\text{CC}}$ the probability of a TOA-TOA-measurement. In our simulations, we set $p_{\text{CC}} = 0.75$.

The calculation of $n_{\text{frames}}$ is straightforward, given by $n_{\text{frames}} = \frac{1}{\text{T} + \tau}$. The key rate per coincidence click, $r_{\text{CC}}$ is also known, since this is the lower-bound we get on the key rate from our simulation. Therefore, only $P_{\text{click}}$ remains to be determined.

To motivate its calculation, we observe that the unjittered probability of exactly one coincidence click per frame, $P_{\text{unjclick}}$ is given as the sum of all the exclusive coincidence click probabilities defined in Section \ref{sec:calcNoiseLossProbs} and calculated according to Appendix \ref{app:prob_calc}, \begin{align*}
    P_{\text{unjclick}} &= p(A\varphi, B\varphi) + p(A\varphi, B\tilde{\varphi}) + p(A\varphi, B\delta) + p(A\varphi, B\xi) + p(A\delta, B\varphi)\\&+ p(A\xi, B\varphi) + p(A\xi, B\delta) + p(A\delta, B\xi) + p(A\delta, B\delta) + p(A\xi, B\xi),
\end{align*} while at the same time $P_{\text{unjclick}}$ can be obtained as the sum of all the entries of $cc'$, where $cc'$ is the click matrix obtained just like $cc$ (see Appendix \ref{app:click_mat_sim}), but with the distinction that it is calculated using the non-normalized matrix $\rho'_1$ from Equ. \eqref{equ:rho1'} instead of the normalized density matrix $\rho_1$. Inspired by this fact, it's easy to see that when taking into account also jitter, the probability for exactly one (jittered) coincidence click per frame can be obtained as the sum of all the elements of $cc'^J$, which is defined as the matrix we get from $cc'$ when applying jitter to it in the same way as layed out in Section \ref{sec:jitter_simulation} for $cc$.

As the entanglement rate is defined as the Entanglement of Formation per second, it can be obtained analogously from the EoF per coincidence click, which means 
\begin{equation*}
    \text{EoF}_{\text{sec}} = \text{EoF}_{\text{cc}} P_{\text{click}} p_{\text{CC}}\, n_{\text{frames}},
\end{equation*}
where $\text{EoF}_{\text{sec}}$ denotes the entanglement rate and $\text{EoF}_{\text{cc}}$ denotes the EoF per coincidence click.

\section{Validation of the use of the inequality constraint in the SDP for lower-bounding the fidelity}\label{app:Schmidt-bound-proof}
As pointed out in Section \ref{sec:lower-bounding_Schmidt}, we have to show that \begin{equation*} \braket{\Psi_{T}|\tilde{\rho}}{\Psi_{T}} \leq \braket{\Psi_{T}|\rho}{\Psi_{T}} \end{equation*}
holds for any $d^2 \times d^2$ - density matrix $\rho$ and any matrix $\tilde{\rho}$ obtained from it by substituting the real part of the first (or, more generally, any) off-diagonal with a lower bound and setting the corresponding imaginary part to an arbitrary value such that $\tilde{\rho}$ is still Hermitian, with $\ket{\Psi_{T}}$ denoting the maximally entangled state $\ket{\Psi_{T}} = \frac{1}{\sqrt{d}}\sum\limits_{j=0}^{d-1}\ket{jj}$. This is equivalent to demonstrating that 
\begin{equation}
    \braket{\Psi_{T}|\Delta \rho}{\Psi_{T}} \geq 0, \, \text{ with } \Delta \rho: = \rho - \tilde{\rho}
\end{equation}

Assume that $\tilde{\rho}$ is obtained from $\rho$ by modifying the $m$-th off-diagonal ($1 \leq m \leq d^2-1$), meaning that for all the pairs $\rho(k, k+m), \rho(k+m, k)$, $k = 1, \dotsc, d^2-m$ their real part is substituted by a lower bound on it. Let $\Delta \rho^{(k)}$ be the matrix obtained by subtracting the matrix where only the elements $\rho(k,k+m)$ and $\rho(k+m,k)$ are modified from the matrix $\rho$, i.e. $\Delta \rho^{(k)}$ is zero everywhere except at the positions $(k,k+m)$ and $(k+m,k)$. Then, 
\begin{equation*}
    \Delta \rho = \sum\limits_{k=1}^{d^2-m} \Delta \rho^{(k)}.
\end{equation*} and therefore, due to linearity of the inner product, we have
\begin{equation*}
    \braket{\Psi_{T}|\Delta \rho}{\Psi_{T}} = \sum\limits_{k=1}^{d^2-m} \braket{\Psi_{T}|\Delta \rho^{(k)}}{\Psi_{T}}.
\end{equation*}
Now let us evaluate $\braket{\Psi_{T}|\Delta \rho^{(k)}}{\Psi_{T}}$ for some fixed $k$.

Let $A:= \text{Re}(\rho)$ be the real part of $\rho$ and $B:= \text{Im}(\rho)$ the imaginary part, such that we have $\rho(n,l) = A(n,l) + iB(n,l)$ and $\rho(l,n) = A(n,l) - iB(n,l)$ for the entries on the upper respectively lower triangular part of $\rho$ and let $\tilde{A}:= \text{Re}(\tilde{\rho})$ and $\tilde{B}:= \text{Im}(\tilde{\rho})$ where $\tilde{A}(n,l) \leq A(n,l)$ by definition of the lower bound. Then we have $\Delta \rho(n,l) = (A(n,l) - \tilde{A}(n,l)) + i(B(n,l) - \tilde{B(n,l)})$ and $\Delta \rho(l,n) = (A(n,l) - \tilde{A}(n,l)) - i(B(n,l) - \tilde{B(n,l)})$.

Focusing on the $m$-th off-diagonal, let $\delta_A^{(k)} := A(k,k+m) - \tilde{A}(k,k+m)$ and $\delta_B^{(k)} := B(k,k+m) - \tilde{B}(k,k+m)$, such that for the upper triangular matrix of $\Delta \rho^{(k)}$ we have $\Delta \rho^{(k)}(k,k+m) = \delta_A^{(k)} + i \delta_B^{(k)}$ and for the lower triangular part $\Delta \rho^{(k)}(k+m,k) = \delta_A^{(k)} - i \delta_B^{(k)}$, where these two entries are the only non-zero entries of $\Delta \rho^{(k)}$.

Therefore, 
\begin{equation*}
    \begin{aligned}              \braket{\Psi_{T}|  \Delta \rho^{(k)}}{\Psi_{T}} &= \bra{\Psi_{T}}\left((\delta_A^{(k)} + i\delta_B^{(k)})\ket{k-1}\bra{k+m-1} + (\delta_A^{(k)} - i\delta_B^{(k)})\ket{k+m-1}\bra{k-1}\right) \ket{\Psi_{T}}\\
        &=(\delta_A^{(k)} + i\delta_B^{(k)}) \braket{\Psi_{T}}{k-1}\braket{k+m-1}{\Psi_{T}} + (\delta_A^{(k)} - i\delta_B^{(k)})\braket{\Psi_{T}}{k+m-1}\braket{k-1}{\Psi_{T}}.
    \end{aligned}
\end{equation*}
Since $\ket{\Psi_{T}}$ is real, it holds that $\braket{\Psi_{T}}{j} = \braket{j}{\Psi_{T}}, \, j = 0, \dotsc, d^2-1$, so that the imaginary parts in above equation cancel out, yielding
\begin{equation*}
    \braket{\Psi_{T}|\Delta \rho}{\Psi_{T}} = \sum\limits_{k=1}^{d^2-m}2\delta_A^{(k)} \braket{k-1}{\Psi_{T}}\braket{k+m-1}{\Psi_{T}}.
\end{equation*}

Defining $w_j:=\braket{j-1,j-1}{\Psi_{T}}$, we have $w_j = \frac{1}{\sqrt{d}}, \, j=1, \dotsc, d$. Furthermore, for all other basis vectors $\ket{n,l}$ with $n\neq l$ the component $\braket{n,l}{\Psi_{T}}$ is equal to zero. Hence the product $\braket{k-1}{\Psi_{T}}\braket{k+m-1}{\Psi_{T}}$ is different from zero only when both factors are of the form $\braket{n,l}{\Psi_{T}}, n=l$, from which it follows that all nonzero terms in above sum can be written as $2 \delta^{(k)} w_i w_j = 2 \delta_A^{(k)} \frac{1}{d}$ (for suitable $i,j$).

Since all the factors are non-negative - the non-negativity of $\delta_A^{k}$ following from its definition -, the sum consists of only non-negative terms and
so we have proven $\braket{\Psi_{T}|\Delta\rho}{\Psi_{T}} \geq 0$, from which our original inequality follows directly.

\section{Comments on the simulations}\label{app:realistic_simulations}
To explore the supposition stated in Section \ref{sec:discussion}, we pursue the following two strategies for our simulation: In both cases, detector efficiency, environmental photon rates, dark count rates and the probability that a detected coincidence click is a TOA-TOA-click are set to experimental realistic values, while 
\begin{itemize}
    \item either the loss probability is fixed realistically and the bin size (which is an experimentally flexible parameter) is optimized to maximize the key rate per second. This optimization is performed over a range of pair production rates, starting from where a positive key rate is first obtained and extending to the point where no positive key rate can be sustained.
    \item or the pair production rate (which is an experimentally tunable parameter) is fixed at a specific value and the bin size is optimized to maximize the key rate per second. In this scenario, the loss probability is varied, starting from zero up to where no positive key rate is obtained anymore.
\end{itemize}

Although the bin size is a tunable parameter, it is physically constrained as it cannot become arbitrarily small. Consequently, a minimal bin size was imposed during the optimization process to reflect these limitations, as detailed below.

\subsection{Jitter restricting the bin size}
Following a nearest-neighbor jitter model, we assume it is negligibly unlikely that jitter causes a detection more than one time-bin away from its actual position. Therefore, to maintain the validity of our whole model, we must restrict the optimization to bin sizes for which for given jitter-$\sigma$ the probability of a click falling outside the immediate neighboring bins is sufficiently small. Note that this is no actual limitation, since for large ratios of jitter-$\sigma$ to bin size, the key rate and EoF are expected to be bad due to the rarity of correct coincidence clicks, as confirmed by the fact that the optimal bin sizes we find are much larger than the minimal bin sizes considered.

Specifically, we only optimize over bin sizes for which the probability of a click being detected outside the correct bin and its two adjacent neighbors is $\leq 5\%$.

To this end, we first determine the minimal bin size compatible with this threshold and then optimize over all bin sizes larger than or equal to it. As extremely small bin sizes are practically infeasible, we impose a minimal bin size of at least $10^{-16}$ seconds in case the calculated one is smaller.

In certain extreme parameter regimes, it can happen that the key rate per second becomes the better the smaller the bin size is, such that the best bin size corresponds to the minimal bin size just described, 
or conversely, the best bin size might appear infinite if the key rate per coincidence click is negative and asymptotically approaches zero. 
However, such cases do not occur within experimentally realistic parameter constellations and thus do not impact the results presented here.

\subsection{Probability of jitter affecting bins beyond the neighboring ones}
As established above, the minimal bin size is constrained to be large enough to uphold the nearest-neighbor jitter assumption, where the minimal bin size is chosen such that the probability of jitter affecting bins beyond the immediate neighbors is at most five percent. However, for the optimized bin sizes determined in our simulations, this probability is found to be several orders of magnitude lower, further confirming that the assumption of nearest-neighbor jitter is legitimate and does not artificially constrain our model.

To illustrate the scale of these probabilities, it shall be mentioned that the maximal observed probability for jitter affecting non-adjacent bins is in the order of $10^{-9}$ (i.e. $10^{-7}\%$) for the simulations with varying pair production rates and in the order of $10^{-10}$ (i.e. $10^{-8}\%$) in the simulations with varying loss probability. These results demonstrate that even for our conservative jitter model (as detailed in Section \ref{subsec:jitter_probs_calc}), in physically relevant parameter regimes, jitter-induced crosstalk beyond adjacent bins is practically negligible.

\section{Details on the results regarding all investigated quantities}\label{app:detailed_results}
Here, details on the observations mentioned in Sections \ref{subsubsec:discuss_EoF_and_others} and \ref{subsubsec:discuss_entRate} are given.
\begin{figure*}[!htbp]
    \centering
    \includegraphics[width=1.\textwidth]{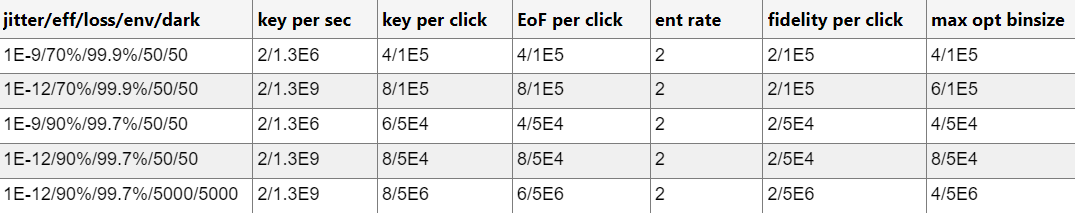}
    \caption{For five different parameter constellations (see column one), the table shows where the maximum of the quantities is achieved (see columns two to seven), with the first number giving the dimension and the second one the corresponding best pair production rate (in emitted photon pairs per second) within this dimension. (e.g. For a jitter-sigma of $10^{-9}$ seconds, detection efficiency of 70\%, photon loss probability of 99.9\% and environmental photon count as well as dark count rate of 50 photons per second, the highest key rate per second - which is the quantity we optimized the bin size for - is obtained for dimension two, at a pair production rate of $1.3*10^6$ photon pairs per second, the highest key rate per coincidence click and EoF per coincidence click both occur at dimension four, at a pair production rate of $10^5$ photons per second, the highest lower bound on the fidelity is largest for dimension two, at a pair production rate of $10^5$ photon pairs per second, and the optimized bin size is largest for dimension four, at a pair production rate of $10^5$ photons per second, while the best entanglement rate is obtained for dimension two. Since it is always attained at the highest pair production rate considered, at which simulations are stopped since the key rate is already negative, this pair production rate is not stated in the table. The last column provides information about the observed minima and maxima of the Schmidt number per dimension two, four, six and eight over pair production rates where the key rate is positive.}
    \label{fig:prod_table}
\end{figure*}

\begin{figure*}[!htbp]
    \centering
    \includegraphics[width=1.\textwidth]{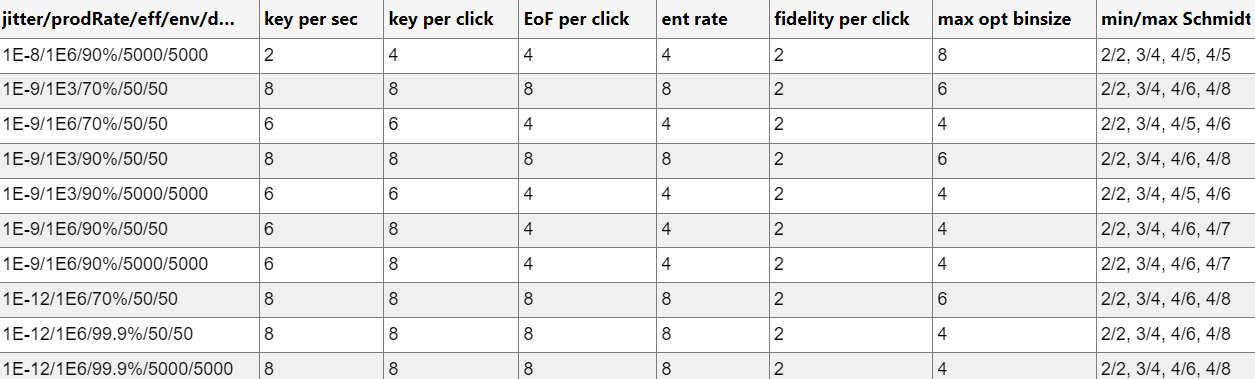}
    \caption{For ten different parameter constellations (see column one: jitter-Sigma/pair production rate/detector efficiency/environmental photon rate/dark count rate), the table shows for which dimension the maximum of the quantities is attained (see column two to seven). Since (except for the maximum over the optimized bin sizes, see the first subplot of Figure \ref{fig:plot_binsizes}) within each dimension the maximum is obtained for zero loss, the parameter loss is not explicitly stated within the table. In the last column, the observed minima and maxima of the Schmidt number per dimension two, four, six and eight from zero loss to a loss level where the key rate ends to be positive, are given.}
    \label{fig:loss_table}
\end{figure*}

\begin{figure*}[!htbp]
    \centering
    \includegraphics[width=1.\textwidth]{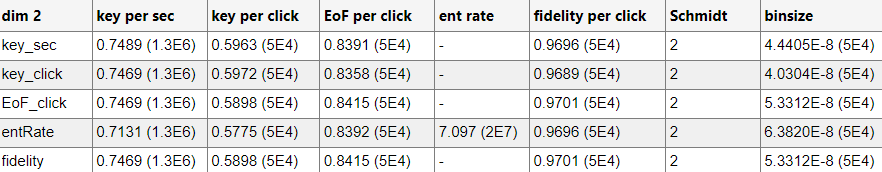}
    \caption{Results for dimension $d = 2$ under a varying pair production rate. The table presents the maxima of the quantities specified column-wise, achieved by optimizing the bin size with respect to the specific target metric defined row-wise. Parentheses indicate the pair production rates at which the respective maxima were observed.}
    \label{fig:prod2_additional_table}
\end{figure*}

\begin{figure*}[!htbp]
    \centering
    \includegraphics[width=1.\textwidth]{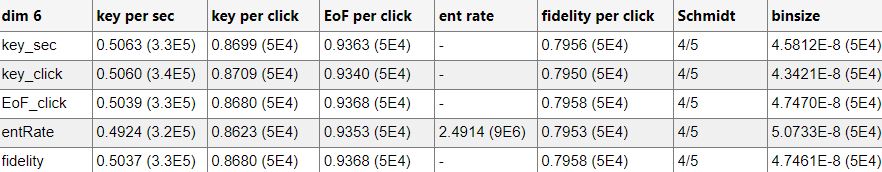}
    \caption{Results for dimension $d = 6$ under a varying pair production rate. The table presents the maxima of the quantities specified column-wise, achieved by optimizing the bin size with respect to the specific target metric defined row-wise. Parentheses indicate the pair production rates at which the respective maxima were observed.}
    \label{fig:prod6_additional_table}
\end{figure*}

\begin{figure*}[!htbp]
    \centering
    \includegraphics[width=1.\textwidth]{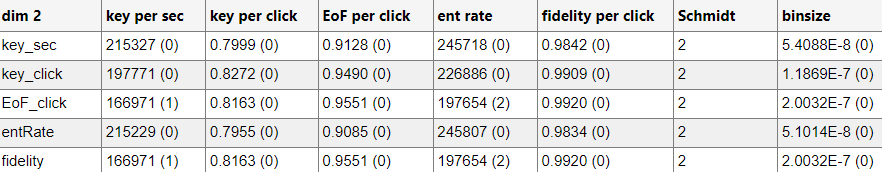}
    \caption{Results for dimension $d = 2$ under a varying loss probability. The table presents the maxima of the column-wise quantities, achieved by optimizing the bin size with respect to the specific target metric defined row-wise. Parentheses indicate the loss probabilities at which the respective maxima were observed.}
    \label{fig:loss2_additional_table}
\end{figure*}

\begin{figure*}[!htbp]
    \centering
    \includegraphics[width=1.\textwidth]{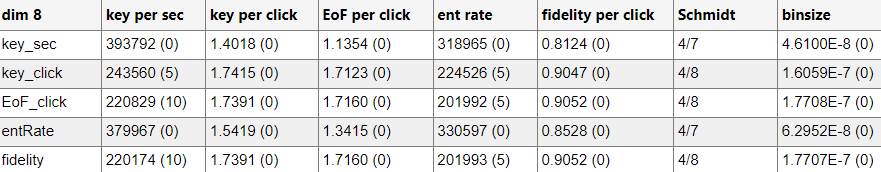}
    \caption{Results for dimension $d = 8$ under a varying loss probability. The table presents the maxima of the column-wise quantities, achieved by optimizing the bin size with respect to the specific target metric defined row-wise. Parentheses indicate the loss probabilities at which the respective maxima were observed.}
    \label{fig:loss8_additional_table}
\end{figure*}

Next, we provide details on some results for the additional optimizations outlined in Section \ref{subsubsec:additional_optim}. Figures \ref{fig:prod2_additional_table} and \ref{fig:prod6_additional_table} illustrate the results for a varying pair production rate, while Figures \ref{fig:loss2_additional_table} and \ref{fig:loss8_additional_table} present the data for a varying loss probability.

In each table, rows two to six of the first column specify the metric for which the bin size was optimized ( key rate per second, key rate per coincidence click, entanglement of formation, entanglement rate and fidelity). The columns display the corresponding maxima for the quantities evaluated column-wise (key rate per second, key rate per click, entanglement of formation, entanglement rate, fidelity, Schmidt number (with both observed minima and maxima provided), and the optimized bin size). The numbers in parentheses indicate the specific pair production rates or loss probabilities at which these maxima were observed.

For all quantities except the entanglement rate, the optimization was performed over the parameter range where a non-negative secret key rate per second was sustained. Since the maximum of the entanglement rate is typically reached at higher pair production rates, its values are omitted in \ref{fig:prod2_additional_table} and \ref{fig:prod6_additional_table} except in the specific row where the entanglement rate itself was the optimization target. In that case, the considered interval for the pair production rate was expanded until the entanglement rate dropped close to zero, enabling the identification of its absolute maximum.

The baseline parameters were fixed to a jitter-sigma of $10^{-9}$ seconds, a detector efficiency of $90\%$, both an environmental photon and a dark count rate of $50$ photons per second. For the scenario of the fixed pair production rate, the rate of emitted photon pairs was set to $10^6$ pairs per second, while in the fixed-loss-scenario, the photon loss probability was set to $99.7\%$. 

Interestingly, for dimensions two and four, each quantity attains its maximum at approximately the same pair production rate, regardless of the target metric used for bin size optimization. In contrast, for dimensions six and eight, the optimal pair production rate for the secret key rate per second exhibits a stronger dependence on the specific optimization metric.

As observed in the data from Figures \ref{fig:loss2_additional_table} and \ref{fig:loss8_additional_table}, optimizing the bin size with respect to per-click metrics can lead to scenarios where per-second quantities reach their maxima at non-zero loss probabilities rather than under ideal transmission conditions. Given that per-click optimization favors much larger bins than optimizing per-second performance, we conjecture that this occurs, first, because higher loss reduces the risk of multi-photon events, thereby enhancing the probability of a valid coincidence click per time frame, and, second, because increasing loss shifts the optimal bin size to smaller values, leading to a higher frame rate. However, as expected, the ideal scenario of zero loss delivers peak results for each quantity when the bin size is specifically optimized for that respective metric. 
\end{document}